\documentclass[twocolumn]{autart}    

\usepackage{amsmath}
\usepackage{amsfonts}
\usepackage{mathtools}
\mathtoolsset{showonlyrefs}
\usepackage{cite}
\usepackage{amssymb}
\usepackage{xcolor}
\usepackage{mathrsfs}
\usepackage[aboveskip=-3pt]{subcaption}
\usepackage{tikz}
\usepackage[title]{appendix}
\makeatletter
\renewcommand*{\@opargbegintheorem}[3]{\trivlist
  \item[\hskip \labelsep{\bfseries #1\ #2}] \textbf{(#3)}\ \itshape}
\usepackage{amsmath}
\usepackage[norelsize,algo2e,ruled,linesnumbered]{algorithm2e}
\usepackage{pgfplots}
\usepgfplotslibrary{groupplots,dateplot}
\usetikzlibrary{patterns,shapes.arrows, fadings}
\usetikzlibrary{automata} 
\usetikzlibrary{positioning} 
\usetikzlibrary{arrows} 
\usetikzlibrary{decorations.text,calc,arrows.meta}
\tikzset{node distance=2.5cm, 
every state/.style={ 
semithick,
fill=gray!10},
initial text={}, 
double distance=2pt, 
every edge/.style={ 
draw,
->,>=stealth', 
auto,
semithick}}
\pgfplotsset{compat=newest}
\newlength\figH
\newlength\figW
\DeclareMathOperator*{\argmax}{arg\,max}
\DeclareMathOperator{\adj}{Adj}

\DeclarePairedDelimiter\ceil{\lceil}{\rceil}

\makeatother
\newcommand{\norm}[1]{\left\lVert#1\right\rVert}
\newcommand{\Var}[1]{\text{Var}\left[#1\right]}
\newcommand{\Expectation}[1]{\mathbb{E}\left[#1\right]}

\newcommand{\Adjacent}[2]{\adj_b(#1, #2)=1}
\newcommand{\rew}{J}
\newcommand{\productthing}{\mu} 
\newcommand{\Rmax}{R_{\text{max}}}
\newcommand{\tRmax}{\tilde{R}_{\text{max}}}

\begin{document}
\begin{frontmatter}

\title{Differentially Private Reward Functions in Policy \\ Synthesis for Markov Decision Processes\thanksref{footnoteinfo}} 

\thanks[footnoteinfo]{This paper was not presented at any IFAC 
meeting. This work was partially supported by NSF under CAREER grant \#1943275, AFRL under grant \#FA8651-23-F-A008, ONR under \#N00014-21-1-2502, and AFOSR under grant \#FA9550-19-1-0169. Any opinions, findings and conclusions or recommendations expressed in this material are those of the authors and do not necessarily reflect the views of sponsoring agencies.}
\author[GT]{Alexander Benvenuti}\ead{abenvenuti3@gatech.edu},    
\author[UF]{Calvin Hawkins}\ead{calvin.hawkins@ufl.edu},               
\author[UF]{Brandon Fallin}\ead{brandonfallin@ufl.edu},  
\author[UF]{Bo Chen}\ead{bo.chen@ufl.edu},  
\author[Eglin]{Brendan Bialy}\ead{brendan.bialy@us.af.mil},  
\author[Eglin]{Miriam Dennis}\ead{miriam.dennis.1@us.af.mil},
\author[GT]{Matthew Hale}\ead{matthale@gatech.edu}
\address[GT]{School of  Electrical and Computer Engineering, Georgia Institute of Technology, Atlanta, GA USA.}  
\address[UF]{Department of  Mechanical and Aerospace Engineering, University of Florida, Gainesville, FL USA.}             
\address[Eglin]{Munitions Directorate, Air Force Research Laboratory, Eglin AFB, FL USA.}        
\begin{keyword}                           
Differential Privacy; Markov Processes; Multi-Agent Systems.               
\end{keyword}                             

\begin{abstract}                          
Markov decision processes often seek to maximize a reward function, but onlookers may infer reward functions by observing the states and actions of such systems, revealing sensitive information. Therefore, in this paper we introduce and compare two methods for privatizing reward functions in policy synthesis for multi-agent Markov decision processes, which generalize Markov decision processes. Reward functions are privatized using differential privacy, a statistical framework for protecting sensitive data. The methods we develop perturb either (1) each agent's individual reward function or (2) the joint reward function shared by all agents. We show that approach (1) provides better performance. We then develop a polynomial-time algorithm for the numerical computation of the performance loss due to privacy on a case-by-case basis. Next, using  approach (1), we develop guidelines for selecting reward function values to preserve ``goal" and ``avoid" states while still remaining private, and we quantify the increase in computational complexity needed to compute policies from privatized rewards. Numerical simulations are performed on three classes of systems and they reveal a surprising compatibility with privacy: using reasonably strong privacy~($\epsilon =1.3$) on average induces as little as a~$5\%$ decrease in total accumulated reward and a~$0.016\%$ increase in computation time.
\end{abstract}
\end{frontmatter}

\section{Introduction}
Many autonomous systems share sensitive information to operate, 
such as sharing locations within teams of autonomous vehicles. As a result, 
there has arisen interest in privatizing such information when it is communicated. 
However, these protections can be difficult to provide in systems in which agents are inherently observable, such as in a traffic system in which agents are visible to other vehicles \cite{glancy2012privacy} or a power system in which power usage is visible to a utility company \cite{guan2018privacy}. These systems do not offer the opportunity to protect information by modifying communications due to the fact to agents are observed directly. 
Thus, a fundamental challenge is that information about agents is visible to observers, though we would still like to limit the inferences that can be drawn from that information.

In this paper, we consider the problem of protecting the reward function of a Markov decision process, even when its states
and actions can be observed. 
In particular, we model individual agents as Markov decision processes (MDPs), and we model collections of agents as multi-agent Markov decision processes (MMDPs). Given that an MDP is simply an MMDP with a single agent, we 
focus our developments on MMDPs in general. 
In MMDPs, agents' goal is to synthesize a reward-maximizing policy. 

Since these agents can be observed, the actions they take could reveal their reward function or some of its properties,
which may be sensitive information. 
For example, quantitative methods can draw such inferences from agents' trajectories both offline \cite{ng2000algorithms, ziebart2008maximum} and online \cite{zhi2020online}. Additionally, onlookers may draw a variety of qualitative inferences from agents as well; see \cite{ramirez2011goal} for more details. Such inferences include relative reward values between state-action pairs or subsets of reward function values. This past work shows that harmful quantitative and qualitative inferences can be drawn
without needing to recover the entire reward function. 
Given the wide variety of inferences that could be drawn, 
we seek to protect agents' reward functions from both existing privacy attacks and those yet to be developed.

We use differential privacy to provide these protections. Differential privacy is a statistical notion of privacy originally used to protect entries of databases \cite{cynthia2006differential}. Differential privacy has been used recently in control systems and filtering \cite{le2013differentially, cortes2016differential, han2018privacy,yazdani2022differentially,hawkins2020differentially}, and to privatize objective functions in distributed optimization \cite{wang2016differentially, huang2015differentially, han2016differentially, nozari2016differentially, dobbe2018customized, lv2020differentially}.
The literature on distributed optimization has already established the principle that agents' objective functions
require privacy, and a key focus of that literature has been convex minimization problems. In this paper, we consider agents maximizing
reward functions and hence are different from that body of work, though the need to protect individual agents' rewards 
is just as essential as the protection of convex objectives in those works. 

Differential privacy is appealing for this purpose because of its strong protections for data and its immunity to post-processing \cite{dwork2014algorithmic}. That is, the outputs of arbitrary computations on differentially private data are still differentially private. 
This property provides protections against observers that make inferences about agents' rewards, including through techniques
that do not exist yet. 
We therefore first privatize reward functions, then use dynamic programming to synthesize a policy with the privatized rewards. Since this dynamic programming stage is post-processing, the resulting policy preserves the privacy of the reward functions, as do observations of an agent executing that policy and any downstream inferences that rely on those observations. 

Of course, we expect perturbations to reward functions to affect performance. To assess the impact of privacy on the agents' performance, we develop an algorithm to quantify the exact sub-optimality of the policy synthesized with the privatized reward function. In particular, it relates the value function (i.e., the total accumulated discounted reward) with privacy to the value function without privacy. We then compute the exact computational complexity of this algorithm in terms of system parameters. These calculations and subsequent simulations both show the tractability of this algorithm. In particular, its computational complexity is shown to be bilinear in the size of the state and action spaces.

Perturbations to rewards also affect the computational complexity of computing policies. The computational complexity of computing policies is a function of the absolute values of the rewards in a tabular representation of the the reward function. As privacy strengthens, their absolute values will increase, and thus, the computational complexity will grow. We analyze the impact of privacy on computational complexity by bounding the average increase in the number of steps of value iteration required to compute a policy with privatized rewards. We then show that privacy induces minimal changes in computational complexity, i.e., we show that the change in computational complexity grows logarithmically with the strength of privacy and in simulation show as little as a~$0.016\%$ increase in computational complexity.

We also provide two types of guidelines for system designers to shape reward functions with privacy in mind. For designing reward values, given a set of state-action pairs with high rewards and another set with low rewards, we provide tight bounds on the probability that these sets maintain their relative ordering in the reward function after privacy is implemented. This insight allows designers to choose reward values that give similar long term behavior while still protecting reward values with privacy. For selecting privacy parameters, we also bound the accuracy of privatized rewards in such a way that system designers may select privacy parameters based on the maximum allowable error in the privatized reward function.

To summarize, we make the following contributions: 
\begin{itemize}
    \item We develop two differential privacy mechanisms for reward functions in MMDPs (Theorems \ref{thm:input} and \ref{thm:output}). 
    \item We provide an analytical bound on the accuracy of privatized reward functions and use that bound to trade-off the strength of privacy and the error it induces in rewards (Theorem \ref{theorem:cheb}).
    \item We provide an algorithm to compute privacy-performance trade-off on a case-by-case basis, then quantify its computational complexity (Theorem \ref{theorem:cop}).
    \item We provide guidelines for designing reward functions to minimize the impact of privacy on reaching or avoiding states of interest (Theorem \ref{theorem:reward_function_design}).
    \item We quantify the increase in computational complexity needed to compute a policy on privatized rewards (Theorem \ref{thm:e_cost_of_privacy}).
    \item We validate the impact of privacy upon performance in three classes of simulations (Section \ref{sec:Waypoint}). 
\end{itemize}

\subsection*{Related Work}
Privacy has previously been considered for Markov decision processes, both in planning~\cite{gohari2020privacy, chen2023differential, venkitasubramaniam2013privacy, hassan2021privacy} and reinforcement learning~\cite{zhou2022differentially, ma2019differentially}. Privacy has also been considered for Markov chains~\cite{fallin2023differential} and for general symbolic systems~\cite{chen2023differentialsymbolic}. The closest work to ours is in~\cite{gohari2020privacy},~\cite{hassan2021privacy}, and~\cite{ye2019differentially}. In~\cite{gohari2020privacy} and~\cite{hassan2021privacy}, privacy is applied to transition probabilities, while we apply
it to reward functions. In~\cite{ye2019differentially}, the authors use differential privacy to protect rewards 
that are learned. 
We differ from this work (and all others on private reinforcement learning) since we consider planning problems with a known reward. 

A preliminary version of this work appeared in~\cite{benvenuti2023differentially}. The current paper adds design guidelines for privacy-amenable reward functions, the average increase in computational cost to compute a policy on privatized rewards, new simulations, and proofs of all results. 

This paper is organized as follows: Section~\ref{sec:background} provides background and problem statements, and Section~\ref{sec:privacy_implementation} presents two methods for privatizing rewards. Section~\ref{sec:accuracy_cheb} formalizes privacy-accuracy trade-offs, and Section~\ref{sec:cost_of_privacy} presents a method of computing the cost of privacy. Section~\ref{sec:reward_design} provides guidelines for designing reward functions with privacy in mind, and Section~\ref{sec:policy_cost} illustrates the trade-off between computational complexity and privacy for computing policies. Section~\ref{sec:Waypoint} provides simulations, and Section~\ref{sec:conclusion} concludes. 

\subsection{Notation}\label{subsec:notation}
For $N\in\mathbb{N}$, we use $[N]$ to denote $\{1, 2, \ldots, N\}$. We use $\Delta (B)$ to be the set of probability distributions over a finite set $B$ and we use $|\cdot|$ for the cardinality of a set. Additionally, we use $\ceil{\cdot}$ as the ceiling function, and~$\log$ to denote the natural logarithm. We use $\pi$ both as the usual constant and as a policy for an MDP since both uses are standard. The meaning will be clear from context. We also use $\Phi(y) = \frac{1}{\sqrt{2\pi}}\int_{-\infty}^{y} e^{-\frac{\theta^2}{2}}d\theta$ to denote the cumulative distribution function (CDF) for the Gaussian distribution and $\mathcal{Q}(y) = 1-\Phi(y)$ to denote the survival function. We also use~$\mathcal{N}(0, \sigma^2 I)$ to denote a multivariate Gaussian distribution with mean vector~$0$ and covariance matrix~$\sigma^2I$, and~$f\circ h$ to be the composition of~$f$ and~$h$.
\section{Preliminaries and Problem Formulation}\label{sec:background}
This section reviews Markov decision processes and differential privacy, then gives formal problem statements. 
\subsection{Markov Decision Processes}\label{subsec:MDP}
Consider a collection of $N$ agents indexed over $i\in[N]$. We model agent $i$ as a Markov decision process.

\begin{defn}[Markov Decision Process]
Consider~$N$ agents. Agent~$i$'s Markov Decision Process (MDP) is the tuple~$\mathcal{M}^i = (\mathcal{S}^i, \mathcal{A}^i, r^i, \mathcal{T}^i)$, where~$\mathcal{S}^i$ and~$\mathcal{A}^i$ are agent $i$'s finite sets of local states and local actions, respectively. Additionally, let~$\mathcal{S} = \mathcal{S}^1\times \cdots\times \mathcal{S}^N$ be the joint state space of all agents. Then~$r^i:\mathcal{S}\times \mathcal{A}^i\rightarrow \mathbb{R}$ is agent~$i$'s reward function, 
 and~$\mathcal{T}^i:\mathcal{S}^i\times \mathcal{A}^i\rightarrow \Delta(\mathcal{S}^i)$ is agent~$i$'s transition probability function. 
\end{defn}

With $\mathcal{T}^i:\mathcal{S}^i\times \mathcal{A}^i\rightarrow \Delta(\mathcal{S}^i)$, we see that $\mathcal{T}^i(s^i, a^i)\in \Delta(\mathcal{S}^i)$ is a probability distribution over the possible next states when taking action $a^i\in\mathcal{A}^i$ in state $s^i\in\mathcal{S}^i$. We say that $\mathcal{T}^i(s^i, a^i, y^i)$ is the probability of transitioning from state $s^i$ to state $y^i \in \mathcal{S}^i$ when agent $i$ takes action $a^i\in \mathcal{A}^i$.
We now model the collection of agents as a multi-agent Markov decision process (MMDP).


\begin{defn}[MMDP; \cite{boutilier1996planning}] \label{def:mmdp}
A multi-agent Markov decision process (MMDP) is the tuple~$\mathcal{M}= (\mathcal{S}, \mathcal{A}, r, \gamma, \mathcal{T})$, 
where~$\mathcal{S} = \mathcal{S}^1 \times \cdots \times \mathcal{S}^N$ is the joint state space,~$\mathcal{A} = \mathcal{A}^1\times \cdots\times \mathcal{A}^N$ is the joint action space,~$r(s, a) = \frac{1}{N}\sum_{i\in[N]} r^i(s, a^i)$ is the joint reward function value for joint state~$s = (s^1, \ldots, s^N)\in\mathcal{S}$ and joint action~$a = (a^1, \ldots, a^N)\in\mathcal{A}$, the constant~$\gamma\in (0, 1]$ is the discount factor on future rewards, and~$\mathcal{T}:\mathcal{S}\times\mathcal{A}\to\Delta(\mathcal{S})$ is the joint transition probability distribution. That is,~$\mathcal{T}(s, a, y)=\prod_{i=1}^N \mathcal{T}^i(s^i, a^i, y^i)$ denotes the probability of transitioning from joint state~$s$ to joint state~$y= (y^1, \ldots, y^N)\in\mathcal{S}$ given joint action~$a$, for all~$s,~y\in\mathcal{S}$ and $a\in\mathcal{A}$.
\end{defn}
\begin{rem}
    An MMDP with $N=1$ reduces to an MDP, and thus results that apply to MMDPs also apply to MDPs. Therefore, we frame the results in this work in terms of MMDPs for generality. 
\end{rem}
 Given a joint action~$a_j\in\mathcal{A}$, agent~$i$ takes the local action~$a^i_{I_j(i)}\in\mathcal{A}^i$, where we use~$I_j(i)$ to denote the index of agent~$i$'s local action corresponding to joint action~$j$. That is, for the action~$a_j\in\mathcal{A}$ we have~$a_j = \big(a^1_{I_j(1)}, a^2_{I_j(2)}, \ldots, a^N_{I_j(N)}\big)$. For 
    $r(s_j, a_k) = \frac{1}{N}\sum_{i\in[N]} r^i(s_j, a^i_{I_k(i)})$,
we define the mapping~$\rew$ such that 
\begin{equation}\label{eq:J}
    r(s_j, a_k)=\rew(\{r^i(s_j, a^i_{I_k(i)})\}_{i\in[N]}).
\end{equation}
\begin{rem}
    We emphasize that the joint reward~$r$ is not simply an average of each agent's rewards. This is a result of agents having differing local action spaces. Such an average over all local actions may not even exist, since agents may have a different number of local actions given a joint state.
    Instead,~$\rew$ iterates over all possible joint states and local actions to produce a tabular representation of the 
    joint reward for the MMDP in a way that includes each agent's rewards when taking each of their possible local 
    actions in each joint state.
\end{rem}

A joint policy~$\pi: \mathcal{S} \rightarrow \mathcal{A}$, represented as~$\pi = (\pi^1, \ldots, \pi^N)$, where~$\pi^i:\mathcal{S}\rightarrow \mathcal{A}^i$ is agent~$i$'s policy for all~$i\in[N]$, is a set of policies which commands agent~$i$ to take action~$\pi^i(s)$
in joint state~$s$. 
The control objective for an MMDP then is: given a joint reward function~$r$, formed by iterating~\eqref{eq:J} over all joint states~$s\in\mathcal{S}$ and joint actions~$a\in\mathcal{A}$, develop a joint policy that solves
\begin{equation}\label{eq:value_iteration}
    \max_{\pi} v_{\pi}(s) = \max_{\pi} E\left[\sum_{t=0}^{\infty} \gamma^t r(s_t, \pi(s_t))\right],
\end{equation}
where we call~$v_{\pi}$ the ``value function", and where~$\Expectation{\cdot}$ denotes the expectation taken over the randomness of agents' state transitions.

Often, it is necessary to evaluate how well a non-optimal policy performs on a given MMDP. To do so, we consider the vectorized form of the value function, denoted~$V_{\pi}$ for a given policy~$\pi$, where the~$i^{\text{th}}$ entry in the vector~$V_{\pi}$, namely~$V_{\pi, i}$, is equal to the value function evaluated at state~$i$,~$v_{\pi}(s_i)$. Accordingly, we state the following proposition that the Bellman operator is a contraction mapping, which we will use in Section \ref{sec:cost_of_privacy} to evaluate any policy on a given MMDP. 
\begin{prop}[Policy Evaluation; \cite{puterman2014markov}]
\label{prop:contract}
    Fix an MMDP~$\mathcal{M} = (\mathcal{S}, \mathcal{A}, r, \gamma, \mathcal{T})$. Let~$V_{\pi}\in\mathbb{R}^{n}$ be the vectorized form of a value function~$v_{\pi}$, let~$V_{\pi, s'}$ be the value function vector at index~$s'$, and let~$\pi$ be a joint policy. Define the Bellman operator $\mathcal{L}: \mathbb{R}^{n}\to \mathbb{R}^{n}$ as
     $\mathcal{L}V_{\pi} \coloneqq \sum_{a\in\mathcal{A}}\pi(s)\left(r(s, a)+\sum_{s'\in\mathcal{S}}\gamma \mathcal{T}(s, a, s')V_{\pi, s'}\right)$.
    Then~$\mathcal{L}$ is a~$\gamma$-contraction mapping with respect to the~$\infty$-norm. 
    That is,~$\norm{\mathcal{L}V_{\pi_1}-\mathcal{L}V_{\pi_2}}_{\infty}\leq \gamma \norm{V_{\pi_1}-V_{\pi_2}}_{\infty}$ for all~$V_{\pi_1}, V_{\pi_2} \in \mathbb{R}^{n}$. 
\end{prop}

Solving an MMDP is~$P$-Complete and is done efficiently via dynamic programming \cite{puterman2014markov}, which we use below. 

\subsection{Differential Privacy}

We now describe the application of differential privacy to vector-valued data and this will be applied to agents' rewards represented as vectors. 
The goal of differential privacy is to make ``similar" pieces of data appear approximately indistinguishable. The notion of ``similar" is defined by an adjacency relation. Many adjacency relations exist, and we present the one used in the remainder of the paper; see \cite{dwork2014algorithmic} for additional background.
\begin{defn}[Adjacency]\label{def:adj2}
    Fix an adjacency parameter~$b>0$ and two vectors~$v, w\in\mathbb{R}^n$. Then~$v$ and~$w$ 
    are adjacent if the following conditions
    hold: (i) There exists some~$j\in[n]$ such that~$v_j \neq w_j$
    and~$v_k = w_k$ for all~$k\in[n]\setminus\{j\}$ and (ii) $\|v - w\|_1 \leq b$,
    where~$\norm{\cdot}_1$ denotes the vector 1-norm. We use~$\Adjacent{v}{w}$ to say $v$ and $w$ are adjacent, and~$\adj_b(v, w)=0$ otherwise.
\end{defn}
\begin{rem}
    Definition \ref{def:adj2} states that $v$ and $w$ are adjacent if they differ in at most one element by at most $b$. Defining adjacency over a single element is common in the privacy literature \cite{dwork2014algorithmic} and does \emph{not} imply that only a single value of a reward function is protected by privacy. Rather, two reward functions with a single differing entry will be made approximately indistinguishable, while reward vectors with $k$ differing entries will still retain a form of approximate indistinguishability. 
    Strong privacy protections have been attained in prior work with notions of adjacency that differ in one entry, and we use this idea here to attain
    those same strong privacy protections. 
    See \cite{dwork2014algorithmic} for more details on adjacency relations.
\end{rem}

Differential privacy is enforced by a randomized map called a ``mechanism.'' 
For a function $f:\mathbb{R}^n\to\mathbb{R}^m$, a mechanism $\mathscr{M}$ approximates $f(x)$ for all inputs~$x$ according to the following definition. 

\begin{defn}[Differential Privacy; \cite{dwork2014algorithmic}]\label{def:dp}
    Fix a probability space $(\Omega, \mathcal{F}, \mathbb{P})$. Let~$b > 0$, $\epsilon>0$, and $\delta\in[0, \frac{1}{2})$ be given. A mechanism $\mathscr{M}:\mathbb{R}^{n}\times\Omega\to\mathbb{R}^{m}$ is ($\epsilon,\delta$)-differentially private if for all $x, x'\in\mathbb{R}^{n}$ that are adjacent in the sense of Definition \ref{def:adj2}, and for all measurable sets $S\subseteq\mathbb{R}^{m}$, we have
            $\mathbb{P}[\mathscr{M}(x)\in S] \leq e^\epsilon \mathbb{P}[\mathscr{M}(x')\in S]+\delta$.
\end{defn}
Definition~\ref{def:dp} gives a quantitative definition of the similarities between the distributions of~$\mathscr{M}(x)$ and ~$\mathscr{M}(x')$ that must be induced by differential privacy. It is from this definition that other favorable properties of differential privacy\textemdash such as immunity to post-processing\textemdash follow from.

In Definition~\ref{def:dp}, the strength of privacy is controlled by the privacy parameters~$\epsilon$ and~$\delta$. In general, smaller values of~$\epsilon$ and~$\delta$ imply stronger privacy guarantees. Here,~$\epsilon$ can be interpreted as quantifying the leakage of sensitive information and~$\delta$ can be interpreted as the probability that differential privacy leaks more information than allowed by~$\epsilon$. Typical values of~$\epsilon$ and~$\delta$ are~$0.1$ to~$10$~\cite{hsu2014differential} and~$0$ to~$0.05$~\cite{hawkins2023node}, respectively. Differential privacy is calibrated using the ``sensitivity" of the function being privatized, which we define next. 
\begin{defn}[Sensitivity]\label{def:sensitivity}
    Fix an adjacency parameter $b>0$. The $\ell_2$-sensitivity of a function $f:\mathbb{R}^n\to\mathbb{R}^m$ is
        $\Delta_2 f = \sup_{x, x': \Adjacent{x}{x'}} \norm{f(x) - f(x')}_2.$     
\end{defn}
In words, the sensitivity encodes how much~$f$ can differ on two adjacent inputs. A larger sensitivity implies that~$f$ can differ more on adjacent inputs, and, to mask these differences, higher variance noise is needed when generating private outputs. We now define a mechanism for enforcing differential privacy, namely the Gaussian mechanism. The Gaussian mechanism adds zero-mean noise drawn from a Gaussian distribution to functions of sensitive data.
\begin{lem}[Gaussian Mechanism; \cite{cynthia2006differential}]\label{lem:gauss_mech}
    Let $b>0$, $\epsilon>0$, and $\delta\in [0, 1/2)$ be given, and fix the adjacency relation from Definition \ref{def:adj2}. The Gaussian mechanism takes sensitive data $f(x)\in\mathbb{R}^m$ as an input and outputs private data
        $\mathcal{G}(x)=f(x)+z$,
    where $z\sim \mathcal{N}(0, \sigma^2I)$. The Gaussian mechanism is $(\epsilon, \delta)$-differentially private if
        ${\sigma\geq \frac{\Delta_2 f}{2\epsilon}\kappa(\epsilon, \delta)}$,
    where~$\kappa(\epsilon, \delta) = \mathcal{Q}^{-1}(\delta)+\sqrt{\mathcal{Q}^{-1}(\delta)^2+2\epsilon}$ and $\mathcal{Q}$ is from Section \ref{subsec:notation}.
\end{lem}
We sometimes consider the identity query $f(x) = x$, which has $\Delta_2 f = b$, where $b$ is from Definition~\ref{def:adj2}. 

\begin{lem}[Immunity to Post-Processing; \cite{dwork2014algorithmic}]\label{lem:arbitrary}
    Let~$\mathscr{M}:\mathbb{R}^n\times \Omega \to \mathbb{R}^m$ be an~$(\epsilon, \delta)$-differentially private mechanism in the sense of Definition~\ref{def:dp}. Let~$h:\mathbb{R}^m\to\mathbb{R}^p$ be an arbitrary mapping. Then~$h\circ\mathscr{M}:\mathbb{R}^n\to\mathbb{R}^p$ is $(\epsilon, \delta)$-differentially private.
\end{lem}

This lemma implies that we can first privatize rewards and then compute a decision policy from those privatized rewards,
and the synthesis and execution of that policy also keep the rewards differentially private because they are post-processing. 


\subsection{Problem Statements}
Consider the policy synthesis problem in~\eqref{eq:value_iteration}. Computing~$\pi^*$ depends on the sensitive reward function~$r^i$ for all~$i\in[N]$. An adversary observing agents execute~$\pi^*$ may attempt to infer~$r^i$ itself or its quantitative or qualitative properties. Therefore, we seek a framework for multi-agent policy synthesis that preserves the privacy of $r^i$ while still performing well. This will be done by solving the following problems.
\begin{prob}\label{Prob:framework}
    Develop privacy mechanisms to privatize individual agents' reward functions in MMDPs. 
\end{prob} 
\begin{prob}\label{prob:accuracy}
    Develop bounds on the accuracy of privatized rewards and use them to formalize trade-offs between privacy and accuracy.
\end{prob}
\begin{prob}\label{prob:cost_of_privacy}
 Develop a computational procedure to evaluate the loss in performance from using a policy generated on the privatized rewards and evaluate its computational complexity.
\end{prob}
\begin{prob}\label{prob:rewarddesign}
 Develop guidelines for designing reward functions for MMDPs that are amenable to privacy. 
\end{prob}
\begin{prob}\label{prob:cop2}
 As a function of privacy parameters, determine the average increase in computational complexity needed to compute a privatized policy. 
\end{prob}

\section{Private Policy Synthesis}\label{sec:privacy_implementation}
In this section, we solve Problem 1. First, we illustrate how we represent reward functions to apply differential privacy. Then, we present two mechanisms for applying privacy to these reward functions. Let~$n_i = |\mathcal{S}^i|$ and~$m_i = |\mathcal{A}^i|$ be the numbers of local states and local actions, respectively, for agent~$i$. The joint state and joint action spaces then have~$n = \prod_{i\in[N]} n_i$ states and~$m = \prod_{i\in[N]} m_i$ actions, respectively.
\subsection{Privacy Setup}
 To use Lemma~\ref{lem:gauss_mech} to enforce differential privacy, we first express the reward function as a vector. We represent the mapping~$r^i$ as a vector~$R^i\in\mathbb{R}^{nm_i}$, where the entries of~$R^i$ correspond to~$r^i$ being evaluated on all of its possible inputs. 
To elaborate, since~$r^i:\mathcal{S}\times\mathcal{A}^i\to\mathbb{R}$, there is a  scalar reward associated with every state-action pair~$(s,a^i)\in\mathcal{S}\times\mathcal{A}^i$ comprised by a joint state~$s$ and local action~$a^i$. Furthermore, agent~$i$ has~$nm_i$ distinct state-action pairs of this kind. 
We thus define~$R^i$ to be the vector with entries~$r^i(s, a^i)$ for all~$s\in\mathcal{S}$ and~$a^i\in\mathcal{A}^i$. 

We use the following
convention for representing~$R^i$.
Denote the joint states by~$s_1, s_2, \ldots, s_{n}$
and denote the local actions
by~$a^i_1, a^i_2, \ldots, a^i_{m_i}$.
Then we set
\begin{multline}\label{eq:big_ri}
R^i = \Big[r^i(s_1, a^i_1), r^i(s_1, a^i_2), \ldots, 
r^i(s_1, a^i_{m_i}),
r^i(s_2, a^i_1),\\
\ldots,
r^i(s_2, a^i_{m_i}), \ldots 
r^i(s_{n}, a^i_1),
\ldots,
r^i(s_{n}, a^i_{m_i})\Big]^T,
\end{multline}
where~$R^i_j$ denotes the~$j^{th}$
entry of the vector~$R^i$.
This vector fixes the joint state~$s_1$ and
computes the reward for this state and each local action.
Then it proceeds to~$s_2$ and does the same for~$s_2$ through~$s_{n}$. This same process can be repeated to represent the agents' 
joint reward~$R\in\mathbb{R}^{nm}$ by fixing the joint state~$s_1$ and computing the reward for this state and each joint action, then doing the same for~$s_2$ through~$s_n$.  

Using Definition~\ref{def:adj2}, we say two reward functions belonging to agent~$i$, denoted~$R^i$ and~$\hat{R}^{i}$, are adjacent if they differ in one entry, with their absolute difference bounded above by~$b$. Let~$R^i$ and~$\hat{R}^i$ correspond to the reward functions~$r^i$ and~$\hat{r}^i$, respectively. Adjacency of~$R^i$ and~$\hat{R}^i$ is equivalent to the existence
of indices~${k \in [n]}$ and~${\ell \in [m_i]}$ such that~$r^i(s_k, a^i_{\ell}) \neq \hat{r}^{i}(s_k, a^i_{\ell})$ and~$r^i(s_c, a^i_d) = \hat{r}^{i}(s_c, a^i_d)$
for all~$c \in [n] \backslash \{k\}$ and~$d \in [m_i] \backslash \{\ell\}$. 

We note that the solution to Problem 1 is not unique, and we consider two means of enforcing privacy. 
In both setups, agents share their reward functions with an aggregator, the aggregator computes
a joint decision policy for the agents, and then the aggregator sends each agent its 
constituent policy from the joint policy. The two setups we consider thus differ in where privacy
is implemented in them. 

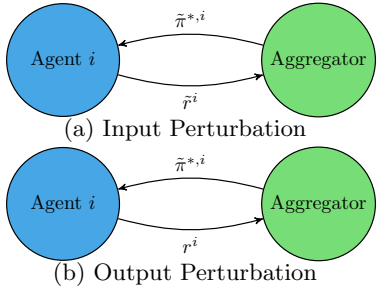
\begin{figure}[h!t]
    \centering
    \begin{subfigure}{0.35\textwidth}
        \centering
         \scalebox{0.75}{
         \definecolor{soft_blue}{RGB}{71, 166, 229}
\definecolor{soft_green}{RGB}{119, 221, 118}
\begin{tikzpicture}[scale = 1]
\node (n1)  [state, minimum size=2cm, fill = soft_blue] {Agent $i$};
\node (n2)  [state,right=of n1, minimum size=2cm, fill = soft_green]   {Aggregator};
\path   (n1)    edge [below, bend right=15] node {$\tilde{r}^i$} (n2)
        (n2)    edge [above, bend right=15] node {$\tilde{\pi}^{*,i}$} (n1);
\end{tikzpicture}}
         \caption{Input Perturbation}
    \label{fig:input_perturbation}
    \end{subfigure}
    \hspace{2cm}
    \begin{subfigure}{0.35\textwidth}
        \centering
    \scalebox{0.75}{
    \definecolor{soft_blue}{RGB}{71, 166, 229}
\definecolor{soft_green}{RGB}{119, 221, 118}
\begin{tikzpicture}[scale = 0.5]
\node (n1)  [state, minimum size=2cm, fill = soft_blue] {Agent $i$};
\node (n2)  [state,right=of n1, minimum size=2cm, fill = soft_green]   {Aggregator};
\path   (n1)    edge [below, bend right=15] node {$r^i$} (n2)
        (n2)    edge [above, bend right=15] node {$\tilde{\pi}^{*,i}$} (n1);
\end{tikzpicture}}
    \caption{Output Perturbation}
    \label{fig:output_perturbation}
    \end{subfigure}
\caption{The flow of information for (a) input perturbation and (b) output perturbation. In input perturbation, 
agent~$i$ sends the aggregator its privatized reward function~$\tilde{r}^i$ (shown by the lower arrow in (a)), while in output perturbation, agent~$i$ sends the sensitive (non-privatized) reward function~$r^i$ (shown by the lower arrow in (b)). By privatizing their rewards before sending them, agents using input perturbation have greater control of the strength of privacy used to protect their rewards, and they do not need to trust the aggregator with their sensitive reward.}
\label{fig:info_flow}
\end{figure}

In the first setup, we apply the Gaussian mechanism to agent~$i$'s vector of rewards,~$R^i$, which is referred to as ``input perturbation". In input perturbation, the aggregator receives a privatized reward, 
denoted~$\tilde{R}^i$ from agent~$i$, for all~$i\in[N]$, and the aggregator uses those privatized policies to generate a policy~$\tilde{\pi}^*$ for the MMDP. 
In the second setup, 
we instead apply the Gaussian mechanism to the vector of joint rewards,~$R$, which is referred to as ``output perturbation". In output perturbation, the aggregator receives sensitive rewards~$r^i$ from agent~$i$ for all~$i\in[N]$, 
computes the vector of joint rewards~$R$,
and applies privacy to it to generate~$\tilde{R}$. 
Then it uses~$\tilde{R}$ to generate a policy~$\tilde{\pi}^* = (\tilde{\pi}^{*,1}, \ldots, \tilde{\pi}^{*,N})$ for the MMDP.
These two approaches are shown in Figure~\ref{fig:info_flow}, and we detail each approach next. 

\subsection{Input Perturbation}
In the case of input perturbation, each agent privatizes the identity map of its own reward. Formally, 
for all~$i \in [N]$
agent~$i$'s reward vector is privatized by taking~$\tilde{R}^i = R^i +w^i$, where~$w^i\sim\mathcal{N}(0, \sigma^2 I)$ for all~$i\in[N]$. The vector~$\tilde{R}^i$ can be put into one-to-one correspondence with a private reward function~$\tilde{r}^i$ in the obvious way. Then, we use~$\tilde{r}(s_j, a_k) = \rew(\{\tilde{r}^i(s_j, a^i_{I_k(i)})\}_{i\in[N]})$ where~$\rew$ is from~\eqref{eq:J} for all~$j\in[n]$ and~$k\in[m]$ to compute~$\tilde{r}$. As a result,~$\tilde{r}$ is the privatized form of the joint reward~$r$ from Definition~\ref{def:mmdp}. The private joint reward~$\tilde{r}$ is then used in place of~$r$ to synthesize the agents' joint policy. After privatization, policy synthesis is post-processing of differentially private data, which implies that the policy also keeps each agent's reward function differentially private due to Lemma~\ref{lem:arbitrary}. Algorithm~\ref{algo:input} presents this method of determining policies from private agent reward functions using input perturbation. 

\begin{algorithm2e}
    \caption{Private Policy Synthesis via Input Perturbation}
    \SetKwFor{ForAll}{for all}{do}{end}
    \label{algo:input}
    \SetAlgoLined
    \textbf{Inputs}: $\{r^i\}_{i\in[N]}$, $\mathcal{S}$, $\mathcal{A}$, $\gamma$, $\mathcal{T}$,
    $\epsilon$, $\delta$, $b$, $N$;\\
    \textbf{Outputs}: Privacy-preserving policies $\{\tilde{\pi}^{*,i}\}_{i\in[N]}$;\\
    \vspace{1mm}
    Agents set $\sigma = \frac{b}{2\epsilon}\kappa(\epsilon, \delta)$;\\
    \ForAll{$i\in[N]$ in parallel}
        {Agent $i$ generates its private reward function with the Gaussian mechanism, $\tilde{R^i}=R^i+w^i$;\\
        Agent $i$ sends $\tilde{R}^i$ to the aggregator;\\}
    Aggregator generates joint reward function: $\tilde{r}(s_j, a_k) = \rew(\{r^i(s_j, a^i_{I_k(i)}\}_{i\in[N]})$ for all $j\in[n]$ and $k\in[m]$;\\
    Aggregator generates joint policy, $\tilde{\pi}^* = \argmax_{\pi} E\left[\sum_{t=0}^{\infty} \gamma^t \tilde{r}(s_t, \pi(s_t))\right]$;\\
    Aggregator sends $\tilde{\pi}^{*,i}$ to agent $i$ for all $i\in[N]$
\end{algorithm2e}

\begin{rem}
    For input perturbation, adjacency in the sense of Definition \ref{def:adj2} implies that every agent's reward may differ in up to one entry by an amount up to~$b$, and these are the differences that must be masked by privacy.
\end{rem}
We next state a theorem proving that input perturbation, i.e., Algorithm~\ref{algo:input}, 
is~$(\epsilon, \delta)$-differentially private.
\begin{thm}[Solution to Problem \ref{Prob:framework}]\label{thm:input}
    Given privacy parameters~$\epsilon>0$,~$\delta\in[0, \frac{1}{2})$, and adjacency parameter~$b>0$, the mapping from~$\{r^i\}_{i\in[N]}$ to~$\{\pi^{*,i}\}_{i\in[N]}$ defined by Algorithm~\ref{algo:input} keeps each~$r^i$~$(\epsilon,\delta)$-differentially private with respect to the adjacency relation in Definition~\ref{def:adj2}.
\end{thm}
\vspace{-0.5cm}
\begin{pf}
    Setting $\sigma= \frac{b}{2\epsilon}\kappa(\epsilon, \delta)$, the result immediately follows from Lemma \ref{lem:gauss_mech} and Lemma \ref{lem:arbitrary} because the computation of $\{\tilde{\pi}^{*,i}\}_{i\in[N]}$ is simply post-processing. \hfill$\blacksquare$
\end{pf}
\vspace{-0.5cm}


Using Algorithm~\ref{algo:input}, each agent enforces the privacy of its own reward function before sending it to the aggregator. Performing input perturbation this way enforces differential privacy on a per-agent basis, which is referred to as ``local differential privacy"~\cite{duchi2013local}. The main advantage of input perturbation is that the aggregator does not need to be trusted since it is only sent privatized information. Another advantage is that agents may select differing levels of privacy. However, to provide a fair comparison with output perturbation, the preceding implementation considers each agent using the same values of~$\epsilon$ and~$\delta$. 
\subsection{Output Perturbation}\label{subsec:output}
In output perturbation, for all~$i \in [N]$ agent~$i$ sends its sensitive (non-private) vector of rewards,~$R^i$,
to the aggregator. 
Then the aggregator uses these rewards to form the joint reward~$R$. For privacy, 
noise is added to the joint reward vector, namely~$\tilde{R} = R + w$, where~$w\sim\mathcal{N}(0, \sigma^2 I)$. Similar to the input perturbation setup, computing the joint policy using the privatized~$\tilde{R}$ is differentially private because computation of the policy is post-processing of private data; this~$\tilde{R}$ can be related to a function~$\tilde{r}$ in the obvious way. Algorithm~\ref{algo:output} presents this method of computing policies when reward functions are privatized using output perturbation, where we use
    $\mu = \max_{j\in[n]}\prod_{\substack{\ell = 1; \ell \neq j}}^N m_\ell. $
\begin{algorithm2e}[t!]
    \caption{Private Policy Synthesis via Output Perturbation}
    \SetKwFor{ForAll}{for all}{do}{end}
    \label{algo:output}
    \SetAlgoLined
    \textbf{Inputs}: $\{r^i\}_{i\in[N]}$, $\mathcal{S}$, $\mathcal{A}$, $\gamma$, $\mathcal{T}$,
        $\epsilon$, $\delta$, $b$, $N$;\\
    \textbf{Outputs}: Privacy-preserving policies $\{\tilde{\pi}^{*,i}\}_{i\in[N]}$;\\
    \vspace{1mm}
    \ForAll{$i\in[N]$ in parallel}
        {Agent $i$ sends reward $r^i$ to the aggregator}
    Aggregator generates joint reward function: $r(s_j, a_k) = \rew(\{r^i(s_j, a^i_{I_k(i)})\}_{i\in[N]})$ for all $j\in[n]$ and $k\in[m]$;\\
    Aggregator sets $\sigma =  \frac{b}{2\epsilon N}\kappa(\epsilon, \delta)\productthing$;\\
    Aggregator generates private reward function with the Gaussian mechanism, $\tilde{R}=R+w$;\\
    Aggregator generates optimal joint policy, $\tilde{\pi}^* = \argmax_{\pi} E\left[\sum_{t=0}^{\infty} \gamma^t \tilde{r}(s_t, \pi(s_t))\right]$;\\
    Aggregator sends $\tilde{\pi}^{*,i}$ to agent $i$ for all $i\in[N]$
\end{algorithm2e}
\begin{rem}\label{remark:output}
    We regard the concatenation of vectorized rewards sent to the aggregator as a single vector, and thus adjacency in Definition 3 says that only a single entry in this vector can change. Thus, adjacency in the setting of output perturbation allows only a single agent's reward to change, and it can change by up to~$b$. However, given the mapping~$\rew$ in~\eqref{eq:J} between all agents' rewards and the joint reward, this difference in one agents' reward will affect the joint reward in several places. We account for this fact next in calibrating the noise required for privacy.
\end{rem}
We next prove a theorem showing that output perturbation, i.e., Algorithm~\ref{algo:output}, is~$(\epsilon, \delta)$-differentially private.

\begin{thm}[Alternative Solution to Problem \ref{Prob:framework}]\label{thm:output}
    Fix privacy parameters~$\epsilon>0$,~$\delta\in[0, \frac{1}{2})$, and adjacency parameter~$b>0$.  
The mapping from~$\{r^i\}_{i\in[N]}$ to~$\{\pi^{*,i}\}_{i\in[N]}$ defined by Algorithm~\ref{algo:output} keeps each~$r^i$~$(\epsilon,\delta)$-differentially private with respect to the adjacency relation in Definition~\ref{def:adj2}.
\end{thm}
\vspace{-0.5cm}
\begin{pf}
See Appendix \ref{apdx:thm_2}.\hfill$\blacksquare$
\end{pf}
\vspace{-0.5cm}

Unlike input perturbation, output perturbation requires that agents trust the aggregator with their sensitive reward functions. Additionally, all agents using output perturbation will have the same level of privacy.

Contrary to the majority other privacy literature, i.e.,~\cite{chen2023differentially, le2013differentially, dwork2014algorithmic}, and references therein, we expect significantly better performance using input perturbation over output perturbation. For output perturbation, the standard deviation~$\sigma$ used to calibrate the noise added for privacy essentially grows exponentially with the number of agents, which is due to the term~$\mu$, 
used in step 7 of Algorithm~\ref{algo:output}. This terms appears because we consider the joint state~$s$ in evaluating~$r^i(s, a^i)$ and a pair of the form~$(s, a^i)\in\mathcal{S}\times \mathcal{A}^i$ will appear many times in~$\Delta_2 \rew$ to account for all possibilities of other agents' actions that can be taken when agent~$i$ takes action~$a^i$. When there are many other agents, many other joint actions are possible, leading to a high sensitivity. In the case of input perturbation, since the joint reward is computed from privatized rewards, the standard deviation of privacy noise does not depend on the number of agents. The standard deviations for varying~$N$ and~$\epsilon$ are shown in Table~\ref{table:the_whole_table}. Table~\ref{table:fix_e_vary_N} shows that the variance of privacy noise is constant for input perturbation for all~$N$, while for output perturbation it grows rapidly with~$N$. Table~\ref{table:fix_N_vary_e} shows that the variance of noise needed for input perturbation is less than that needed for output perturbation for the same level of privacy, quantified by~$\epsilon$, when~$N>1$.
\begin{table}
\begin{subtable}{0.45\columnwidth}
\centering
\begin{tabular}{ccc}
\hline
$N$&Algo. \ref{algo:input}&Algo. \ref{algo:output}\\
\hline
$1$ & $2.524$ &$2.524$\\
$2$ & $2.524$ &$5.049$\\
$5$ & $2.524$ &$129.3$\\
$10$ & $2.524$ &$66,174$\\
\hline
\end{tabular}
\caption{}
\label{table:fix_e_vary_N}
\end{subtable}
\hspace{0.5cm}
\begin{subtable}{0.45\columnwidth}
\centering
\begin{tabular}{ccc}
\hline
$\epsilon$&Algo. \ref{algo:input}&Algo. \ref{algo:output}\\
\hline
$0.1$ & 23.48 &46.95\\
$1$ & 2.524 &5.049\\
$5$ & 0.6251 &1.250\\
$10$ & 0.3684 &0.7367\\
\hline
\end{tabular}
\caption{}
\label{table:fix_N_vary_e}
\end{subtable}
\caption{Standard Deviation required to provide differential privacy with (a)~$\epsilon=1$ for varying~$N$ and (b)~$N=2$ for varying~$\epsilon$. In (a), we see that the standard deviation of privacy noise is constant for input perturbation for all~$N$, while for output perturbation it grows rapidly with~$N$. In (b), the standard deviation of noise needed for input perturbation is less than that needed for output perturbation for every~$\epsilon$.}
\label{table:the_whole_table}
\end{table}
\begin{rem}\label{remark:input_supremacy}
    Given the significantly lower standard deviation of privacy noise needed for input perturbation versus output perturbation, we focus on input perturbation in the remainder of the paper, unless stated otherwise.
\end{rem}
The difference in performance between input and output perturbation is explored further in Examples~\ref{ex:toy} and~\ref{example:gridworld} in Section~\ref{sec:Waypoint}. In the next section, we analyze 
and quantify privacy-accuracy trade-offs.
\section{Accuracy Analysis}\label{sec:accuracy_cheb}
In this section, we solve Problem \ref{prob:accuracy}. Specifically, we analyze the accuracy of the reward functions that are privatized using input perturbation. To do so, we compute an upper bound on the expected maximum absolute error between the sensitive, non-private joint reward $r$ and the privatized reward~$\tilde{r}$, namely~$\Expectation{\max_{k,j}\left|\tilde{r}(s_k, a_j)-r(s_k, a_j)\right|}$. This bound provides a worst-case bound on expected error for all entries of a reward. Then we use this accuracy bound to develop guidelines for calibrating privacy, i.e., 
guidelines for
choosing~$\epsilon$ given a bound on the 
maximum error,~$A$. The maximum absolute error is bounded as follows.

\begin{thm}[Solution to Problem \ref{prob:accuracy}]\label{theorem:cheb}
    Fix privacy parameters~$\epsilon>0$,~$\delta\in[0, \frac{1}{2})$, adjacency parameter~$b>0$, and MMDP~$\mathcal{M} = (\mathcal{S}, \mathcal{A}, r, \gamma, \mathcal{T})$ with~$N$ agents,~$n$ joint states, and~$m$ joint actions. Let the privatized reward~$\tilde{r}$ be defined as in Algorithm~\ref{algo:input} and let~$\kappa$ be defined as in Lemma~\ref{lem:gauss_mech}. 
    Then
    \begin{equation}\label{eq:accuracy_result}
        \mathbb{E}[\max_{k,j}\left|\tilde{r}(s_k, a_j)-r(s_k, a_j)\right|]\leq
        \frac{Cb}{2\epsilon}\kappa(\epsilon, \delta),
    \end{equation}
    where $C = \sqrt{\frac{2}{N\pi}}
       +\sqrt{\left(1-\frac{2}{\pi}\right) \frac{(nm-1)}{N}}$.
\end{thm}
\vspace{-0.5cm}
\begin{pf}
See Appendix~\ref{apx:thm_3}.\hfill$\blacksquare$
\end{pf}
\vspace{-0.5cm}

\begin{figure}
    \centering
    \begin{subfigure}{0.4\textwidth}
    
        \centering
        \hspace*{-1.5cm}
%
%
\definecolor{chocolate2267451}{RGB}{226,74,51}
\definecolor{dimgray85}{RGB}{85,85,85}
\definecolor{gainsboro229}{RGB}{229,229,229}
\definecolor{lightgray204}{RGB}{204,204,204}
\definecolor{steelblue52138189}{RGB}{52,138,189}
\definecolor{black}{RGB}{0, 0, 0}
\definecolor{GTblue}{RGB}{0, 48, 87}
\definecolor{GTgold}{RGB}{179, 163, 105}
\definecolor{UFOrange}{RGB}{250, 70, 22}
\definecolor{UFblue}{RGB}{0, 33, 165}
\begin{tikzpicture}

\begin{axis}[%
width=0.35\figW,
height=0.5\figH,
axis background/.style={fill=gainsboro229},
axis line style={white},
scale only axis,
xlabel=\textcolor{black}{{Strength of Privacy, $\epsilon$}},
xtick style={color=dimgray85},
x grid style={white},
yminorticks=true,
y grid style={white},
ylabel=\textcolor{black}{Maximum Error},
xmajorgrids,
ymajorgrids,
yminorgrids,
legend cell align={left},
legend style={fill opacity=0.8, draw opacity=1, text opacity=1, draw=white, fill=white},
legend pos = north east,
tick align=outside,
tick pos=left,
]
\addplot [color=UFOrange, ultra thick]
  table[row sep=crcr]{%
0.1	39.7208603958748\\
0.2	20.0374217902452\\
0.3	13.4742622395102\\
0.4	10.1912472082821\\
0.5	8.22034501197768\\
0.6	6.90554153933937\\
0.7	5.96568516279309\\
0.8	5.26019792930417\\
0.9	4.71097926310048\\
1	4.27116742571574\\
1.1	3.91094016333505\\
1.2	3.61041501766432\\
1.3	3.35582640623424\\
1.4	3.13734108871024\\
1.5	2.94774744008537\\
1.6	2.78163624104876\\
1.7	2.6348705967474\\
1.8	2.50423254563981\\
1.9	2.38718126251221\\
2	2.2816837980438\\
2.1	2.18609417635333\\
2.2	2.09906546412961\\
2.3	2.01948477670941\\
2.4	1.94642453132126\\
2.5	1.87910539842402\\
2.6	1.81686780176705\\
2.7	1.75914975093322\\
2.8	1.70546942332564\\
2.9	1.65541134925058\\
3	1.60861535943346\\
3.1	1.56476767124404\\
3.2	1.52359364583202\\
3.3	1.48485186177552\\
3.4	1.44832923422865\\
3.5	1.41383697049678\\
3.6	1.38120719942579\\
3.7	1.35029014714865\\
3.8	1.32095175856351\\
3.9	1.29307168455694\\
4	1.26654157098167\\
4.1	1.24126359788246\\
4.2	1.21714922727373\\
4.3	1.19411812552869\\
4.4	1.17209723260972\\
4.5	1.1510199553062\\
4.6	1.130825465616\\
4.7	1.11145808861734\\
4.8	1.09286676678601\\
4.9	1.0750045898422\\
5	1.05782838095731\\
5.1	1.04129833158923\\
5.2	1.02537767840348\\
5.3	1.0100324167251\\
5.4	0.995231045788726\\
5.5	0.980944341742573\\
5.6	0.967145154939431\\
5.7	0.953808228534331\\
5.8	0.94091003581935\\
5.9	0.928428634074322\\
6	0.916343533008253\\
6.1	0.904635576118597\\
6.2	0.893286833511293\\
6.3	0.882280504909391\\
6.4	0.871600831737036\\
6.5	0.861233017302499\\
6.6	0.851163154222234\\
6.7	0.841378158330291\\
6.8	0.831865708406296\\
6.9	0.822614191132438\\
7	0.813612650757238\\
7.1	0.804850743002679\\
7.2	0.796318692802719\\
7.3	0.788007255506328\\
7.4	0.779907681217802\\
7.5	0.772011681982008\\
7.6	0.764311401552928\\
7.7	0.756799387511051\\
7.8	0.749468565519169\\
7.9	0.742312215527443\\
8	0.735323949757464\\
8.1	0.728497692311866\\
8.2	0.721827660270987\\
8.3	0.715308346151408\\
8.4	0.708934501613103\\
8.5	0.702701122312599\\
8.6	0.696603433809048\\
8.7	0.690636878438688\\
8.8	0.684797103080831\\
8.9	0.679079947745419\\
9	0.673481434918393\\
9.1	0.667997759606731\\
9.2	0.662625280030041\\
9.3	0.657360508910165\\
9.4	0.65220010531439\\
9.5	0.64714086701156\\
9.6	0.642179723303807\\
9.7	0.63731372829966\\
9.8	0.632540054597106\\
9.9	0.627855987347691\\
10	0.623258918675058\\
};
\addlegendentry{Theorem~\ref{theorem:cheb}}
\addplot [color=UFblue, dashed, ultra thick]
  table[row sep=crcr]{%
0.1	20.9275149693127\\
0.2	10.5499742385307\\
0.3	7.10998148389198\\
0.4	5.3772122990243\\
0.5	4.32387884455732\\
0.6	3.64003450841965\\
0.7	3.14161418311892\\
0.8	2.76862958587682\\
0.9	2.48903071663763\\
1	2.24575689392769\\
1.1	2.05945297901443\\
1.2	1.90064060889723\\
1.3	1.77199571450253\\
1.4	1.64882863965127\\
1.5	1.55553830572708\\
1.6	1.47162932570219\\
1.7	1.39170153452806\\
1.8	1.31233764225138\\
1.9	1.25958876378651\\
2	1.1981370697873\\
2.1	1.15365778345425\\
2.2	1.10828119353643\\
2.3	1.06113320630613\\
2.4	1.02474417811314\\
2.5	0.989396810965822\\
2.6	0.958273139843883\\
2.7	0.927215437291377\\
2.8	0.897551991314673\\
2.9	0.874091373968244\\
3	0.848266192069361\\
3.1	0.828204993174522\\
3.2	0.802789663126033\\
3.3	0.785078151405782\\
3.4	0.761327572401597\\
3.5	0.739827243572178\\
3.6	0.727668906443031\\
3.7	0.712319964636376\\
3.8	0.694942145191678\\
3.9	0.677448655431695\\
4	0.667457123679098\\
4.1	0.654586513490895\\
4.2	0.63944618431853\\
4.3	0.626717677189308\\
4.4	0.615718946089812\\
4.5	0.607992569783985\\
4.6	0.5955173352967\\
4.7	0.586009402737604\\
4.8	0.575487259346837\\
4.9	0.568515466618461\\
5	0.558674558946811\\
5.1	0.547822616238816\\
5.2	0.537431475408637\\
5.3	0.529694265813408\\
5.4	0.525067752562715\\
5.5	0.51462893531921\\
5.6	0.50975153965427\\
5.7	0.502512173841918\\
5.8	0.492110742124444\\
5.9	0.486417324015006\\
6	0.484299842039031\\
6.1	0.473886474186053\\
6.2	0.468302043604107\\
6.3	0.465894280100989\\
6.4	0.458298240734996\\
6.5	0.453579087794988\\
6.6	0.449229087839582\\
6.7	0.442603257068292\\
6.8	0.437264306317729\\
6.9	0.433991337037564\\
7	0.428756335710666\\
7.1	0.423498415145481\\
7.2	0.42020902798524\\
7.3	0.414547126820445\\
7.4	0.411482001602542\\
7.5	0.405782550772308\\
7.6	0.403917150434088\\
7.7	0.396440979509874\\
7.8	0.395300772257108\\
7.9	0.391613430039181\\
8	0.386868179364387\\
8.1	0.384175001982135\\
8.2	0.378613383775093\\
8.3	0.376979718158389\\
8.4	0.372266171371487\\
8.5	0.370846514839033\\
8.6	0.366684901949226\\
8.7	0.365098099640248\\
8.8	0.359872948850848\\
8.9	0.359213052722855\\
9	0.355825398898525\\
9.1	0.352984762685371\\
9.2	0.349437792389054\\
9.3	0.345659290084845\\
9.4	0.343129530045302\\
9.5	0.341674846070128\\
9.6	0.339452208662678\\
9.7	0.334068641526893\\
9.8	0.332993699185446\\
9.9	0.330388962018758\\
10	0.32736379928487\\
};
\addlegendentry{True}

\end{axis}
\end{tikzpicture}%
             \caption{}
    \label{fig:cheb_1norm}
    \end{subfigure}

    \begin{subfigure}{0.4\textwidth}
   
    \centering
         \hspace*{-1.5cm}
%
%
\definecolor{chocolate2267451}{RGB}{226,74,51}
\definecolor{dimgray85}{RGB}{85,85,85}
\definecolor{gainsboro229}{RGB}{229,229,229}
\definecolor{lightgray204}{RGB}{204,204,204}
\definecolor{steelblue52138189}{RGB}{52,138,189}
\definecolor{black}{RGB}{0, 0, 0}
\definecolor{GTblue}{RGB}{0, 48, 87}
\definecolor{GTgold}{RGB}{179, 163, 105}
\definecolor{UFOrange}{RGB}{250, 70, 22}
\definecolor{UFblue}{RGB}{0, 33, 165}
\begin{tikzpicture}

\begin{axis}[%
width=0.35\figW,
height=0.4\figH,
axis background/.style={fill=gainsboro229},
axis line style={white},
scale only axis,
xlabel=\textcolor{black}{{Maximum Allowable Average Error, $A$}},
xtick style={color=dimgray85},
x grid style={white},
yminorticks=true,
y grid style={white},
ylabel=\textcolor{black}{Minimum $\epsilon$},
xmajorgrids,
ymajorgrids,
yminorgrids,
tick align=outside,
tick pos=left,
]
\addplot [color=UFOrange, ultra thick]
  table[row sep=crcr]{%
1	5.36738895389554\\
1.1	4.76115212095136\\
1.2	4.27402720750469\\
1.3	3.87467709542152\\
1.4	3.54173949169284\\
1.5	3.26018422941048\\
1.6	3.01914829243328\\
1.7	2.81059679622311\\
1.8	2.62846600417914\\
1.9	2.46809832853075\\
2	2.32586001961291\\
2.1	2.19887646052318\\
2.2	2.08484510400059\\
2.3	1.98190083474829\\
2.4	1.88851745282536\\
2.5	1.80343450251675\\
2.6	1.72560218242386\\
2.7	1.65413935056802\\
2.8	1.58830114516537\\
2.9	1.52745375630822\\
3	1.47105457811197\\
3.1	1.41863645322199\\
3.2	1.36979506132021\\
3.3	1.32417874567259\\
3.4	1.28148024678461\\
3.5	1.24142994002333\\
3.6	1.20379026834424\\
3.7	1.16835113149722\\
3.8	1.13492604589007\\
3.9	1.10334892932993\\
4	1.07347139547274\\
4.1	1.04516046639021\\
4.2	1.0182966299589\\
4.3	0.992772183068178\\
4.4	0.968489812881519\\
4.5	0.94536137727614\\
4.6	0.923306852660559\\
4.7	0.902253423030921\\
4.8	0.882134688680931\\
4.9	0.862889976662401\\
5	0.844463738084795\\
5.1	0.826805019783755\\
5.2	0.809866999890203\\
5.3	0.793606578479412\\
5.4	0.777984015841643\\
5.5	0.762962612046212\\
5.6	0.748508422412421\\
5.7	0.734590004287754\\
5.8	0.721178191193958\\
5.9	0.708245890957206\\
6	0.695767904907665\\
6.1	0.683720765631055\\
6.2	0.672082591092278\\
6.3	0.660832953238669\\
6.4	0.649952759435996\\
6.5	0.639424145300686\\
6.6	0.629230377672397\\
6.7	0.619355766626536\\
6.8	0.609785585560571\\
6.9	0.600505998504021\\
7	0.591503993902695\\
7.1	0.582767324215188\\
7.2	0.574284450735788\\
7.3	0.566044493124392\\
7.4	0.558037183182149\\
7.5	0.550252822462401\\
7.6	0.542682243351206\\
7.7	0.535316773290965\\
7.8	0.528148201855307\\
7.9	0.521168750413916\\
8	0.514371044152939\\
8.1	0.507748086240569\\
8.2	0.501293233948533\\
8.3	0.495000176559079\\
8.4	0.488862914903778\\
8.5	0.482875742395379\\
8.6	0.477033227427282\\
8.7	0.471330197027049\\
8.8	0.465761721661066\\
8.9	0.46032310109694\\
9	0.455009851238817\\
9.1	0.449817691858469\\
9.2	0.444742535151884\\
9.3	0.439780475057359\\
9.4	0.434927777276671\\
9.5	0.430180869945953\\
9.6	0.425536334907529\\
9.7	0.420990899538032\\
9.8	0.416541429091931\\
9.9	0.412184919522954\\
10	0.407918490749003\\
};
\end{axis}
\end{tikzpicture}%
    \caption{}
    \label{fig:cheb_1norm2}
    \end{subfigure}
    \caption{Simulation of the bounds in (a) Theorem~\ref{theorem:cheb} and (b) Corollary~\ref{corollary:choose_epsilon} with~$\delta = 0.01$,~$nm = 8$, and~$b = 1$. In (a), we see that Theorem~\ref{theorem:cheb} captures the qualitative behavior of the empirically computed expected maximal error, while also providing tight bounds on the true maximal error. In (b), increasing privacy strength (decreasing~$\epsilon$) from~$\epsilon=3$ to~$\epsilon = 2$ leads to negligible increases in maximum average error. Furthermore, the increase in maximum average error is small until~$\epsilon\leq2,$ indicating marginal performance losses until privacy is relatively strong. 
    } 
\end{figure}
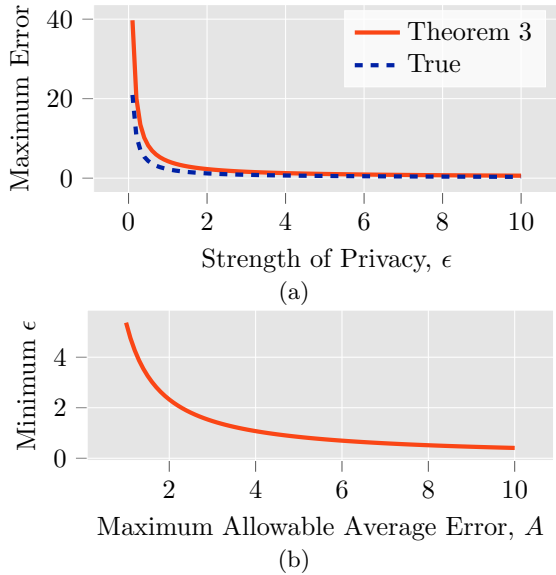

\begin{cor}\label{corollary:choose_epsilon}
    Fix~$\delta\in[0, \frac{1}{2})$ and let the conditions from Theorem~\ref{theorem:cheb} hold. Then a sufficient condition to achieve~$\Expectation{\max_{k,j}|\tilde{r}(s_k, a_j)-r(s_k, a_j)|}\leq A$ is given by
        $\epsilon\geq \frac{2C^2b^2}{4A^2}+\frac{Cb\mathcal{Q}^{-1}(\delta)}{A},$
    where~$C = \sqrt{\frac{2}{N\pi}}
       +\sqrt{\left(1-\frac{2}{\pi}\right) \frac{(nm-1)}{N}}$, and~$\mathcal{Q}$ is defined in Section~\ref{subsec:notation}.
\end{cor}
\vspace{-0.5cm}
\begin{pf}
See Appendix \ref{apdx:cor1}.\hfill$\blacksquare$
\end{pf}
\vspace{-0.5cm}
Theorem~\ref{theorem:cheb} shows that the accuracy of the privatized rewards depends on the number of state-action pairs in the MMDP, which is~$nm$, the number of agents~$N$, and the privacy parameters~$\epsilon$,~$\delta$, and~$b$. The term~$C$ captures the effect of the size of the MMDP, and as a result, the bound is~$\mathcal{O}(\sqrt{nm})$, indicating relatively slow growth with increasing MMDP size. This bound is illustrated in Figure~\ref{fig:cheb_1norm}. We see that as privacy weakens, this bound becomes tighter and it is tight in the the reasonably strong privacy regime~$(\text{e.g. for }\epsilon\in[1, 3])$. This tightness allows users to select their privacy parameters in terms of this worst case absolute error before runtime.

Corollary~\ref{corollary:choose_epsilon} provides a trade-off between privacy and accuracy that allows users to design the strength of privacy,~$\epsilon$, according to some expected maximum absolute error,~$A$. Errors small in magnitude, i.e., of magnitude~$10^0$, allow for reasonably strong privacy~$(\text{e.g. for }\epsilon\in[1, 3])$, highlighting the compatibility between accuracy and privacy in MMDPs. Figure~\ref{fig:cheb_1norm2} illustrates this bound for an example MMDP with~$\delta = 0.01$, $nm = 8$, and~$b = 1$. In the next section we discuss an algorithm for computing the long term performance loss due to privacy in terms of the value function associated with an MMDP.

\section{Cost of Privacy}\label{sec:cost_of_privacy}
In this section, we solve Problem~\ref{prob:cost_of_privacy}. 
To do so, we compare (i) 
an optimal policy generated on the agents' original, non-private reward functions, denoted~$\pi^*$, 
and (ii) 
an optimal policy generated on the agents' privatized reward functions, denoted~$\tilde{\pi}^*$.
Beginning at some joint state~$s_0 \in \mathcal{S}$, the value function~$v_{\pi^*}(s_0)$ encodes the performance of the MMDP given~$\pi^*$ and~$v_{\tilde{\pi}^*}(s_0)$ encodes the performance of the MMDP given~$\tilde{\pi}^*$. We analyze the loss in performance due to privacy, and we do so using the ``cost of privacy" metric introduced in~\cite{gohari2020privacy}, namely~$|v_{\tilde{\pi}^*}(s_0) - v_{\pi^*}(s_0)|$.
Thus, we must compute~$v_{\pi^*}(s_0)$ and~$v_{\tilde{\pi}^*}(s_0)$ to quantify the cost of privacy. Note that there is not a closed form for computing a value function in general. However, Proposition~\ref{prop:contract} provides a method for empirically evaluating the value function for a policy on a given MMDP, namely by repeatedly applying the Bellman operator.
This procedure would need to be implemented on a case-by-case basis to find a value function, and next we state a theorem on the number of operations required to compute the cost of privacy in this way.
\begin{thm}[Solution to Problem \ref{prob:cost_of_privacy}]\label{theorem:cop}
    Fix privacy parameters~$\epsilon>0$ and ~$\delta\in[0, \frac{1}{2})$ and an MMDP~$\mathcal{M} = (\mathcal{S}, \mathcal{A}, r, \gamma, \mathcal{T})$. Let~$\tilde{r}$ be the output of Algorithm~\ref{algo:input}. Then the number of computations needed to compute the cost of privacy~$|v_{\tilde{\pi}^*}(s_0) - v_{\pi^*}(s_0)|$ to within~$\eta$ of its exact value is 
        $nm(K_1+K_2),$ 
    where,
       $ K_1 = \ceil{\log(\frac{4R_{\text{max}}}{\eta(1-\gamma)^2})/\log(\frac{1}{\gamma})},$ $K_2 = \ceil{\log(\frac{4\tilde{R}_{\text{max}}}{\eta(1-\gamma)^2})/\log(\frac{1}{\gamma})}$, and ~$R_{\text{max}} = \max_{s, a}|r(s, a)|$,~$\tilde{R}_{\text{max}} = \max_{s, a}|\tilde{r}(s, a)|$.
\end{thm}
\vspace{-0.5cm}
\begin{pf} 
See Appendix~\ref{adpx:thm_4}. \hfill$\blacksquare$
\end{pf}
\vspace{-0.5cm}
In words,~$nm$ is the number of computations required within each iteration of policy evaluation, while the term in the parentheses is the number of iterations of policy evaluation required. This provides users with an algorithm to compute the loss in performance before runtime for a given private reward~$\tilde{r}$. As privacy strengthens,~$\tRmax$ will grow, indicating the time needed to compute the cost of privacy will increase logarithmically, and thus privacy may be strengthened with minimal impact to the time needed to compute the cost of privacy. 

If~$\pi^*$ is generated using value iteration, the user already has access to~$V_{\pi^*}$ from the policy synthesis process and thus does not need to compute it again for the cost of privacy. In that case, only~$V_{\tilde{\pi}^*}$ needs to be computed. However, we state the total number of computations required assuming the user only has access to~$\pi^*$ and~$\tilde{\pi}^*$. In the next section, we provide guidelines for designing reward functions that are amenable to privacy, in the sense that ``goal" state-action pairs may still be reached and ``bad" state-action pairs can be avoided with a desired probability while still obfuscating reward values.
\section{Privacy-aware Reward Function Design}\label{sec:reward_design}
In this section, we solve Problem~\ref{prob:rewarddesign}. In some privacy literature, the sensitive quantity cannot be designed, such as dynamic trajectories~\cite{le2013differentially} and social relationships~\cite{chen2023differentially}, and we must privatize such data as given. However, in the case of MDPs we may have a system designer who can design both the agent's reward functions and calibrate privacy, allowing them to shape the reward functions with privacy in mind. For example, in some problems the set of absorbing state-action pairs of the MMDP may change without much consequence, such as when there are many acceptable goal state-action pairs~\cite{keizer2013training}, while in other problems, a set of state-action pairs may need to remain the same, e.g., if remaining in a certain set of state-actions pairs is safety-critical~\cite{wachi2020safe}. Similarly, some applications require the set of state-action pairs with the lowest reward value to remain the same after privatization, as they can encode hazards or obstacles that inhibit goal satisfaction~\cite{chen2023differential}. To this end, we consider the scenario where a system designer has both a set of~$p\in\mathbb{N}$ state-action pairs with the highest value rewards and a set of~$q\in\mathbb{N}$ state-action pairs with the lowest rewards values, and they desire for these sets to remain the highest~$p$ and lowest~$q$ after privatization, respectively. 

We now introduce notation to mathematically define the events that the collections of the~$p$ largest and~$q$ smallest entries are not changed by privacy.
Given~$p,q\in\mathbb{N}$, let~$T^p(R^i)$ be the vector of the~$p$ largest rewards in~$R^i$ and let~$T^{-p}(R^i)$ contain the other elements of~$R^i$. Similarly let~$\tilde{T}^p(R^i)$ and~$\tilde{T}^{-p}(R^i)$ be their privatized forms.  Additionally, let~$T_q(R^i)$ be the vector of the~$q$ smallest rewards in~$R^i$ and let~$T_{-q}(R^i)$ contain the other elements of~$R^i$. Let~$T_q(\tilde{R}^i)$ and~$T_{-q}(\tilde{R}^i)$ be their privatized forms. Let~$p^{-} = nm-p$ and~$q^{-} = nm-q$. In computing these sets, all ties are broken arbitrarily.
We now formally define the event that the largest~$p$ entries of~$R^i$ remain the largest~$p$ entries of~$R^i$ after privatization as
\begin{equation}\label{eq:max_rewards_remain}
        B^p = \bigg\{\mathcal{M}(R^i) \,\,\Bigg|\!
        \bigcap_{\substack{\ell\in [p], k\in[p^{-}]}} \!\!\!\!\!\!\! \left[\tilde{T}^p(R^i)_{\ell}\geq \tilde{T}^{-p}(R^i)_{k}\right]\bigg\},
\end{equation}
    regardless of ordering, and the event that the smallest~$q$ entries remain the smallest~$q$  after privatization as
\begin{equation}\label{eq:min_rewards_remain}
        B_q = \bigg\{\mathcal{M}(R^i)\,\,\Bigg|\bigcap_{\substack{\ell\in [q], k\in[q^{-}]}} \!\!\!\!\!\!\!\left[\tilde{T}_q(R^i)_{\ell}\leq \tilde{T}_{-q}(R^i)_{k}\right]\bigg\},
\end{equation}
regardless of ordering, where~$\tilde{T}^p(R^i)_{\ell}$ is the~$\ell^{\text{th}}$ element of~$\tilde{T}^p(R^i)$ and~$\tilde{T}_q(R^i)_\ell$ is the~$\ell^{\text{th}}$ element of~$\tilde{T}_q(R^i)$.

We now state our main results on~$B^p$ and~$B_q.$
Specifically, we will bound the probability of the event~$B^p\cap B_q.$
This event corresponds to when the both the~$p$ largest rewards and~$q$ smallest rewards remain the~$p$ largest and~$q$ smallest after privatization, respectively.
\begin{thm}[Solution to Problem~\ref{prob:rewarddesign}]\label{theorem:reward_function_design}
    Fix privacy parameters~$\epsilon>0$, $\delta\in[0, \frac{1}{2})$, adjacency parameter~$b>0$, and MMDP~$\mathcal{M} = (\mathcal{S}, \mathcal{A}, r, \mathcal{T})$. Let~$\tilde{r}^i$ be defined as in Algorithm~\ref{algo:input}.
    With the events $B^p$ and $B_q$ defined in~\eqref{eq:max_rewards_remain} and~\eqref{eq:min_rewards_remain}, we have
    \begin{multline}
        \!\!\!\!\mathbb{P}(B^p \cap B_q) \leq \min\Big\{\Phi\Big(\frac{\min{T^p(R^i)} - \max{T^{-p}(R^i)}}{\sqrt{2}\sigma} \Big),\\ \mathcal{Q}\Big(\frac{\max{T_q(R^i)} - \min{T_{-q}(R^i)}}{\sqrt{2}\sigma} \Big)\Big\},
    \end{multline}
    where~$\sigma = \frac{b}{2\epsilon}\kappa(\epsilon, \delta)$, and~$\mathcal{Q}$ and~$\Phi$ are from Section~\ref{subsec:notation}.
\end{thm}
\vspace{-0.5cm}
\begin{pf}
See Appendix \ref{adpx:thm_5}. \hfill$\blacksquare$
\end{pf}
\vspace{-0.5cm}
Theorem~\ref{theorem:reward_function_design} shows that the probability of preserving ``goal" and ``avoid" state-action pairs depends the the smallest elements in $T^p(R^i)$ and~$T_{-q}(R^i)$ and the largest elements in~$T^{-p}(R^i)$ and~$T_q(R^i)$, as well as privacy parameters~$\epsilon$, $\delta$, and~$b$. Specifically, as the difference between the elements most likely to swap position after privatization increases, the probability of the elements remaining in the same position increases. This is intuitive because 
a larger difference between elements is less likely to be outweighed by privacy noise.

\begin{figure}
\centering
\begin{tikzpicture}

\definecolor{chocolate2267451}{RGB}{226,74,51}
\definecolor{dimgray85}{RGB}{85,85,85}
\definecolor{gainsboro229}{RGB}{229,229,229}
\definecolor{lightgray204}{RGB}{204,204,204}
\definecolor{steelblue52138189}{RGB}{52,138,189}
\definecolor{black}{RGB}{0, 0, 0}
\definecolor{GTblue}{RGB}{0, 48, 87}
\definecolor{GTgold}{RGB}{179, 163, 105}
\definecolor{UFOrange}{RGB}{250, 70, 22}
\definecolor{UFblue}{RGB}{0, 33, 165}

\begin{axis}[
axis background/.style={fill=gainsboro229},
axis line style={white},
height=0.7\figH,
legend cell align={left},
legend style={fill opacity=0.8, draw opacity=1, text opacity=1, draw=white, fill=white},
legend pos=south east,
tick align=outside,
tick pos=left,
width=0.45\figW,
x grid style={white},
xlabel=\textcolor{black}{$|\min T^p(R^i) - \max T^{-p}(R^i)|$},
xmajorgrids,
xtick style={color=dimgray85},
y grid style={white},
ylabel=\textcolor{black}{$\mathbb{P}(B^p\cap B_q)$}, ymajorgrids,
ytick style={color=black}
]
\addplot [ultra thick, UFOrange]
table[row sep=crcr] {%
0.1	0.528185988898508\\
0.3	0.583997985713682\\
0.5	0.638163195084118\\
0.7	0.689691026781155\\
0.9	0.737740859893462\\
1.1	0.781661683162554\\
1.3	0.821014663677836\\
1.5	0.855577816826758\\
1.7	0.885334028804176\\
1.9	0.910445403636639\\
2.1	0.931218053045048\\
2.3	0.948061921468457\\
2.5	0.961450064128229\\
2.7	0.971881098061362\\
2.9	0.97984751281873\\
3.1	0.985811366631006\\
3.3	0.99018779251168\\
3.5	0.993335835609591\\
3.7	0.995555515028043\\
3.9	0.997089666796095\\
4.1	0.998129048022228\\
4.3	0.998819303518663\\
4.5	0.999268641706659\\
4.7	0.999555366483934\\
4.9	0.999734709943874\\
5.1	0.99984466982868\\
5.3	0.999910756123988\\
5.5	0.99994968903894\\
5.7	0.999972171860019\\
5.9	0.999984898478968\\
6.1	0.99999196008712\\
6.3	0.999995800894342\\
6.5	0.999997848610268\\
6.7	0.99999891876159\\
6.9	0.999999466974103\\
7.1	0.999999742258088\\
7.3	0.999999877758675\\
7.5	0.999999943136372\\
7.7	0.999999974056853\\
7.9	0.999999988391634\\
};
\addlegendentry{Theorem~\ref{theorem:reward_function_design}}
\addplot [ultra thick, dashed, UFblue]
table[row sep=crcr] {%
0.1	0.36104\\
0.3	0.421494214942149\\
0.5	0.48298\\
0.7	0.54259\\
0.9	0.604192083841677\\
1.1	0.661906619066191\\
1.3	0.717677176771768\\
1.5	0.76719\\
1.7	0.80832\\
1.9	0.847906958139163\\
2.1	0.88187\\
2.3	0.90765\\
2.5	0.93091\\
2.7	0.94841\\
2.9	0.96347\\
3.1	0.97296\\
3.3	0.98106\\
3.5	0.98661\\
3.7	0.99157\\
3.9	0.99457\\
4.1	0.99659\\
4.3	0.9978\\
4.5	0.99842\\
4.7	0.99896\\
4.9	0.99931\\
5.1	0.99966\\
5.3	0.99986\\
5.5	0.99991\\
5.7	0.9999\\
5.9	0.99995\\
6.1	0.99999\\
6.3	1\\
6.5	1\\
6.7	1\\
6.9	1\\
7.1	1\\
7.3	1\\
7.5	1\\
7.7	1\\
7.9	1\\
};
\addlegendentry{True}
\end{axis}

\end{tikzpicture}
   \caption{The probability that the ``goal" state remains the same after the implementation of privacy as a function of the difference between reward values. Due to the symmetry between $\Phi(y)$ and $\mathcal{Q}(y)$, this bound is identical for a decreasing gap between the smallest and second smallest rewards. This bound maintains the same qualitative behavior as the true value, and thus allows users to control the probability that their reward functions have the same ``goal" and ``avoid" state-action pairs after privacy is applied.}
   \label{fig:prob_varying_max}
\end{figure}
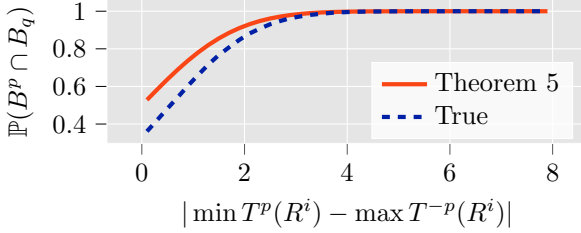

Theorem~\ref{theorem:reward_function_design} is illustrated in Figure~\ref{fig:prob_varying_max}. 
This bound allows users to control which state-action pairs agents will attempt to reach and avoid even after privacy is implemented based on the numerical values in their chosen reward function. Specifically, if users desire the ``goal" and ``avoid" states to change after privacy is implemented, then Theorem~\ref{theorem:reward_function_design} provides a threshold for what numerical reward values make that change more likely, and if they would like them to stay the same, what numerical reward values make that change less likely.

\section{Expected Computational Cost of Privacy}\label{sec:policy_cost}
\begin{figure}[h]
\centering

\begin{subfigure}{0.4\textwidth}
\centering
\hspace*{-1cm}
%
%
\definecolor{chocolate2267451}{RGB}{226,74,51}
\definecolor{dimgray85}{RGB}{85,85,85}
\definecolor{gainsboro229}{RGB}{229,229,229}
\definecolor{lightgray204}{RGB}{204,204,204}
\definecolor{steelblue52138189}{RGB}{52,138,189}
\definecolor{GTblue}{RGB}{0, 48, 87}
\definecolor{GTgold}{RGB}{179, 163, 105}
\definecolor{UFOrange}{RGB}{250, 70, 22}
\definecolor{UFblue}{RGB}{0, 33, 165}
\begin{tikzpicture}

\begin{axis}[%
axis background/.style={fill=gainsboro229},
axis line style={white},
height=0.66\figH,
legend cell align={left},
legend style={fill opacity=0.8, draw opacity=1, text opacity=1, draw=white, fill=white},
legend style={nodes={scale=0.8, transform shape}},
tick align=outside,
tick pos=left,
width=0.4\figW,
x grid style={white},
xlabel=\textcolor{black}{Strength of Privacy, $\epsilon$},
xmajorgrids,
xtick style={color=dimgray85},
ytick style={color=dimgray85},
y grid style={white},
ylabel=\textcolor{black}{$\Expectation{C}$},
xmode=log,
xmajorgrids,
xminorgrids,
ymajorgrids,
]
\addplot [color=UFOrange, ultra thick]
  table[row sep=crcr]{%
0.01	16718.3520534643\\
0.0115139539932645	16377.3460904312\\
0.0132571136559011	16036.5689426673\\
0.0152641796717523	15696.054470207\\
0.0175751062485479	15355.8414105176\\
0.0202358964772516	15015.9740388247\\
0.0232995181051537	14676.5029044648\\
0.0268269579527973	14337.4856477307\\
0.0308884359647748	13998.9879004126\\
0.0355648030622313	13661.0842713379\\
0.0409491506238043	13323.8594154947\\
0.0471486636345739	12987.4091815967\\
0.0542867543932386	12651.8418280188\\
0.0625055192527397	12317.2792907229\\
0.0719685673001152	11983.858478935\\
0.0828642772854684	11651.7325648838\\
0.0954095476349994	11321.0722229219\\
0.109854114198756	10992.0667611025\\
0.12648552168553	10664.9250753486\\
0.145634847750124	10339.8763436555\\
0.167683293681101	10017.1703666874\\
0.193069772888325	9697.07745347085\\
0.222299648252619	9379.88774886848\\
0.255954792269954	9065.90990553683\\
0.294705170255181	8755.46901942823\\
0.339322177189533	8448.90377628134\\
0.390693993705462	8146.56279748038\\
0.449843266896945	7848.80022591182\\
0.517947467923121	7555.97065259094\\
0.596362331659464	7268.42354720112\\
0.6866488450043	6986.49741280578\\
0.79060432109077	6710.51392859315\\
0.910298177991522	6440.7723670306\\
1.04811313415469	6177.54456800577\\
1.20679264063933	5921.07072088576\\
1.38949549437314	5671.55614875269\\
1.59985871960606	5429.1692142284\\
1.84206996932672	5194.04038290805\\
2.12095088792019	4966.26239900732\\
2.44205309454865	4745.89145782551\\
2.81176869797423	4532.94920781889\\
3.23745754281764	4327.42538476882\\
3.72759372031494	4129.28087156007\\
4.29193426012878	3938.45098652416\\
4.94171336132383	3754.84882650001\\
5.6898660290183	3578.36852245999\\
6.55128556859551	3408.88830078766\\
7.54312006335461	3246.27327801881\\
8.68511373751353	3090.37794819145\\
10	2941.04834818127\\
};
\addlegendentry{Theorem~\ref{thm:e_cost_of_privacy}}

\addplot [color=UFblue, dashed, ultra thick]
  table[row sep=crcr]{%
0.01	13901.6562684866\\
0.0115139539932645	13526.026450596\\
0.0132571136559011	13225.8379911148\\
0.0152641796717523	12869.6072853613\\
0.0175751062485479	12539.3176923105\\
0.0202358964772516	12205.1683497763\\
0.0232995181051537	11858.9829646521\\
0.0268269579527973	11510.2480266648\\
0.0308884359647748	11182.5176311828\\
0.0355648030622313	10840.186603361\\
0.0409491506238043	10499.3664828803\\
0.0471486636345739	10150.1584343247\\
0.0542867543932386	9836.93391675006\\
0.0625055192527397	9478.85433391873\\
0.0719685673001152	9149.54797507481\\
0.0828642772854684	8820.03899845278\\
0.0954095476349994	8492.69966520152\\
0.109854114198756	8167.63177182427\\
0.12648552168553	7823.57927157682\\
0.145634847750124	7513.73909041372\\
0.167683293681101	7188.92895941922\\
0.193069772888325	6887.57897132051\\
0.222299648252619	6559.29799589802\\
0.255954792269954	6259.97284265568\\
0.294705170255181	5942.59610448227\\
0.339322177189533	5657.41591407544\\
0.390693993705462	5395.95928891222\\
0.449843266896945	5113.24154439952\\
0.517947467923121	4854.60234811049\\
0.596362331659464	4582.00209112934\\
0.6866488450043	4341.96113862756\\
0.79060432109077	4092.3080452885\\
0.910298177991522	3875.05343116277\\
1.04811313415469	3658.2168380766\\
1.20679264063933	3449.94016233934\\
1.38949549437314	3243.95171493404\\
1.59985871960606	3046.63303743918\\
1.84206996932672	2877.75351484221\\
2.12095088792019	2698.24222089455\\
2.44205309454865	2543.65948759656\\
2.81176869797423	2381.80209718983\\
3.23745754281764	2252.80681902083\\
3.72759372031494	2108.70555002631\\
4.29193426012878	1971.4838657772\\
4.94171336132383	1877.09322161971\\
5.6898660290183	1749.40141315746\\
6.55128556859551	1657.64198266408\\
7.54312006335461	1558.82296550188\\
8.68511373751353	1469.13629621387\\
10	1388.51648038958\\
};
\addlegendentry{True}
\end{axis}

\begin{axis}
[%
axis y line*=right,
axis x line=none,
axis line style={white},
height=0.6\figH,
legend cell align={left},
width=0.4\figW,
x grid style={white},
xlabel=\textcolor{black}{Privacy Strength, $\epsilon$},
xmajorgrids,
xtick style={color=dimgray85},
ytick style={color=dimgray85},
y grid style={white},
yticklabel={$\pgfmathprintnumber{\tick}$\%},
tick align=outside,
xmin=0.001,
xmax=10,
ymin=0,
ymax=25,
]
\end{axis}
\end{tikzpicture}%
    \caption{$R_{\text{max}} = 1$}
\end{subfigure}
\begin{subfigure}{0.4\textwidth}
\centering
\hspace*{-1cm}
%
%
\definecolor{chocolate2267451}{RGB}{226,74,51}
\definecolor{dimgray85}{RGB}{85,85,85}
\definecolor{gainsboro229}{RGB}{229,229,229}
\definecolor{lightgray204}{RGB}{204,204,204}
\definecolor{steelblue52138189}{RGB}{52,138,189}
\definecolor{GTblue}{RGB}{0, 48, 87}
\definecolor{GTgold}{RGB}{179, 163, 105}
\definecolor{UFOrange}{RGB}{250, 70, 22}
\definecolor{UFblue}{RGB}{0, 33, 165}
\begin{tikzpicture}

\begin{axis}[%
axis background/.style={fill=gainsboro229},
axis line style={white},
height=0.66\figH,
legend cell align={left},
legend style={fill opacity=0.8, draw opacity=1, text opacity=1, draw=white, fill=white},
legend style={nodes={scale=0.8, transform shape}},
tick align=outside,
tick pos=left,
width=0.4\figW,
x grid style={white},
xlabel=\textcolor{black}{Strength of Privacy, $\epsilon$},
xmajorgrids,
xtick style={color=dimgray85},
y grid style={white},
ylabel=\textcolor{black}{$\Expectation{C}$},
xmode=log,
xmajorgrids,
xminorgrids,
ymajorgrids,
]
\addplot [color=UFOrange, ultra thick]
  table[row sep=crcr]{%
0.01	11148.476639373\\
0.0115139539932645	10811.1909992438\\
0.0132571136559011	10474.6846147055\\
0.0152641796717523	10139.0706096547\\
0.0175751062485479	9804.47771044166\\
0.0202358964772516	9471.05214019417\\
0.0232995181051537	9138.95966181334\\
0.0268269579527973	8808.38775410029\\
0.0308884359647748	8479.54789398608\\
0.0355648030622313	8152.67790307674\\
0.0409491506238043	7828.04429827863\\
0.0471486636345739	7505.94456388345\\
0.0542867543932386	7186.70923619891\\
0.0625055192527397	6870.70366212214\\
0.0719685673001152	6558.3292611664\\
0.0828642772854684	6250.02408850966\\
0.0954095476349994	5946.26246796083\\
0.109854114198756	5647.55344300668\\
0.12648552168553	5354.43778733844\\
0.145634847750124	5067.48333054558\\
0.167683293681101	4787.27839744678\\
0.193069772888325	4514.42323735341\\
0.222299648252619	4249.51943632536\\
0.255954792269954	3993.15746029314\\
0.294705170255181	3745.90266202356\\
0.339322177189533	3508.28028429009\\
0.390693993705462	3280.76018113696\\
0.449843266896945	3063.7421288547\\
0.517947467923121	2857.54267722823\\
0.596362331659464	2662.38447438893\\
0.6866488450043	2478.38887247891\\
0.79060432109077	2305.57239160415\\
0.910298177991522	2143.84731106504\\
1.04811313415469	1993.02631059181\\
1.20679264063933	1852.83074943451\\
1.38949549437314	1722.9018945615\\
1.59985871960606	1602.81422571495\\
1.84206996932672	1492.08987098849\\
2.12095088792019	1390.21325840544\\
2.44205309454865	1296.64518675998\\
2.81176869797423	1210.835692999\\
3.23745754281764	1132.23529156141\\
3.72759372031494	1060.30435508422\\
4.29193426012878	994.520575140436\\
4.94171336132383	934.384574492178\\
5.6898660290183	879.423835098882\\
6.55128556859551	829.195161242467\\
7.54312006335461	783.285920726579\\
8.68511373751353	741.3143068422\\
10	702.92884731081\\
};
\addlegendentry{Theorem~\ref{thm:e_cost_of_privacy}}

\addplot [color=UFblue, dashed, ultra thick]
  table[row sep=crcr]{%
0.01	8324.92740712756\\
0.0115139539932645	7967.9183276629\\
0.0132571136559011	7651.6395114282\\
0.0152641796717523	7321.10568585222\\
0.0175751062485479	6982.50891216921\\
0.0202358964772516	6677.62672278611\\
0.0232995181051537	6328.21696798152\\
0.0268269579527973	6022.11096500545\\
0.0308884359647748	5713.79636295656\\
0.0355648030622313	5393.25196122956\\
0.0409491506238043	5095.03887251962\\
0.0471486636345739	4793.03009169406\\
0.0542867543932386	4509.94688140868\\
0.0625055192527397	4247.30185228366\\
0.0719685673001152	3989.29102474987\\
0.0828642772854684	3708.57564195178\\
0.0954095476349994	3454.95849340203\\
0.109854114198756	3229.56062215199\\
0.12648552168553	2988.37349087809\\
0.145634847750124	2785.2293277608\\
0.167683293681101	2554.5160879801\\
0.193069772888325	2379.0667233128\\
0.222299648252619	2183.41697367064\\
0.255954792269954	2017.23340625453\\
0.294705170255181	1864.00347357287\\
0.339322177189533	1715.31696570572\\
0.390693993705462	1575.15407403029\\
0.449843266896945	1450.83390386712\\
0.517947467923121	1342.15286170219\\
0.596362331659464	1235.13896052025\\
0.6866488450043	1140.82157727903\\
0.79060432109077	1057.42666528799\\
0.910298177991522	971.174136844623\\
1.04811313415469	902.676818681136\\
1.20679264063933	841.414083270548\\
1.38949549437314	784.73079622381\\
1.59985871960606	733.703474245911\\
1.84206996932672	686.730997189082\\
2.12095088792019	644.656508496668\\
2.44205309454865	613.119068160425\\
2.81176869797423	576.033964506394\\
3.23745754281764	546.350086578012\\
3.72759372031494	521.153386801085\\
4.29193426012878	495.811010390259\\
4.94171336132383	474.314606568222\\
5.6898660290183	455.557959150628\\
6.55128556859551	438.925631823903\\
7.54312006335461	422.821439418618\\
8.68511373751353	408.900771956006\\
10	396.164351967465\\
};
\addlegendentry{True}

\end{axis}

\begin{axis}
[%
axis y line*=right,
axis x line=none,
axis line style={white},
height=0.6\figH,
legend cell align={left},
width=0.4\figW,
x grid style={white},
xlabel=\textcolor{black}{Privacy Strength, $\epsilon$},
xmajorgrids,
xtick style={color=dimgray85},
tick align=outside,
ytick style={color=dimgray85},
y grid style={white},
yticklabel={$\pgfmathprintnumber{\tick}$\%},
xmin=0.01,
xmax=10,
ymin=0,
ymax=15,
]
\end{axis}
\end{tikzpicture}%
    \caption{$R_{\text{max}} = 10$}
\end{subfigure}
   \caption{The expected added computational complexity of computing a policy on a privatized reward function with (a) $\Rmax = 1$ and (b) $\Rmax = 10$, along with $\delta = 0.1$, $b = 1$, $n = 16$, $m = 16$, $\eta = 10^{-8}$, $\gamma = 0.99$, and a range of~$\epsilon$'s. In both cases, one reward is set to~$\Rmax$ and all others are set to~$-\Rmax$. The bound is accurate over the whole range of~$\epsilon$ and maintains the qualitative behavior of the true values, indicating that Theorem~\ref{thm:e_cost_of_privacy} provides accurate estimates for the increase in computation time that privacy induces. With a larger maximum absolute reward value, there is only a minor increase in computation time to compute a policy with the private reward compared to computing a policy with the sensitive reward. Even with strong privacy, such as $\epsilon  = 1$, we observe less than a $10\%$ increase in computation time in practice, indicating that strong privacy protections can be provided without significantly increasing computation time. }
   \label{fig:c_diff}
\end{figure}
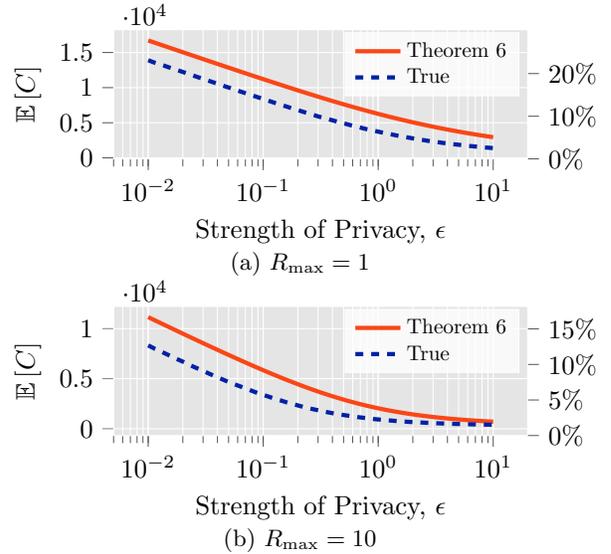

In this section, we solve Problem~\ref{prob:cop2}. In addition to affecting the performance of agents, implementing privacy may also change the computation time needed to compute a policy. In Section~\ref{sec:cost_of_privacy}, we analyzed the cost of privacy on a case-by-case basis. Following a similar construction to Theorem~\ref{theorem:cop}, the number of iterations needed to compute both a policy on the sensitive rewards and a policy on the privatized rewards using value iteration is
    $nm(K_1+K_2),$ 
where~$R_{\text{max}} = \max_{s, a}|r(s, a)|$ and~$\tilde{R}_{\text{max}} = \max_{s, a}|\tilde{r}(s, a)|$, and each value function is computed to within~$\frac{\eta}{2}$ of its exact value. The left term in the parentheses is the number of iterations of a deterministic policy and the right term in the parentheses is the number of iterations of a privately generated policy. We then define
    $C = nm(K_1 - K_2)$
as the difference in the number iterations that are required to compute the privately generated policy and the deterministic policy. We expect~$\Expectation{C}$ to grow with stronger privacy because, given the same initial value functions, the largest absolute value of each entry of the value function will grow with strengthening privacy, and thus more iterations of value iteration will be needed to converge to the optimal value function. 
To solve Problem~\ref{prob:cop2}, we bound~$\Expectation{C}$.
\begin{thm}[Solution to Problem~\ref{prob:cop2}]\label{thm:e_cost_of_privacy}
    Fix privacy parameters~$\epsilon>0$ and~$\delta\in[0, \frac{1}{2})$. Let $\tilde{r}$ be the output of Algorithm \ref{algo:input}. Given an MMDP~$\mathcal{M} = (\mathcal{S}, \mathcal{A}, r, \gamma, \mathcal{T})$, and letting $C = nm(K_1 - K_2)$, the expected difference in the number of iterations need to compute~$\tilde{\pi}^*$ and~$\pi^*$ when each of their value functions are computed to within $\frac{\eta}{2}$ of their exact value is 
    \begin{equation}
        \Expectation{C}\! \leq \!nm\!\!\left(\ceil*{\frac{\log\left(\frac{4R_{\text{max}}
       +4\sigma\sqrt{\left(1-\frac{2}{\pi}\right) \frac{(nm-1)}{N}}}{\eta(1-\gamma)^2}\right)}{\log\left(\frac{1}{\gamma}\right)}}\!\!+\!1 \!- \!K_2\!\!\right),
    \end{equation}
    where~$\sigma = \frac{b}{2\epsilon}\kappa(\epsilon, \delta)$, ~$\Rmax = \max_{s, a}|r(s, a)|$, and~$K_2$ is defined in Theorem~\ref{theorem:cop}.
\end{thm}
\vspace{-0.5cm}
\begin{pf}
See Appendix~\ref{apdx:thm_6}. \hfill$\blacksquare$
\end{pf}
\vspace{-0.5cm}
Note that the bound in Theorem~\ref{thm:e_cost_of_privacy} is~$\mathcal{O}\left(\log\left(1/\sqrt{\epsilon}\right)\right)$. This result shows that it remains tractable to compute policies on privatized rewards while significantly strengthening privacy, as shown by the the logarithmic growth rate of the computational complexity. Theorem~\ref{thm:e_cost_of_privacy} is illustrated in Figure~\ref{fig:c_diff} as (i) the exact increase in iterations and (ii) a percent increase in iterations. We see that this bound remains tight regardless of privacy strength. Furthermore, with realistically chosen values of~$\Rmax$, computing a reward on a privatized policy is not only tractable, but incurs only a minor increase in computation time, even with strong privacy.


\section{Numerical Simulations}\label{sec:Waypoint}
In this section, we consider two multi-agent examples and a single agent example to numerically simulate the preceding theoretical results. First, in Section~\ref{ex:toy}, we consider a team of~$N$ agents, each with the MDP shown in Figure~\ref{fig:toy_MDP}. This example will be used to discuss how the cost of privacy, expressed as a percent change in the value function, i.e,
    $|v_{\tilde{\pi}^*}(s_0) - v_{\pi^*}(s_0)|/|v_{\pi^*}(s_0)|\cdot 100\%,$
scales with increasing numbers of agents. Second, in Section~\ref{example:gridworld}, we will consider a~$4\times 4$ gridworld example to discuss the cost of privacy with a fixed number of agents and varying strengths of privacy,~$\epsilon$. Third, in Section~\ref{ex:waypoint}, we consider a larger MDP with a single agent to assess how differential privacy changes a state trajectory. 
\subsection{Example 1: Simple MDP}
\label{ex:toy}
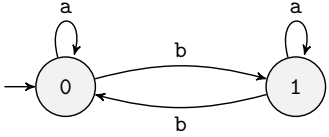
\begin{figure}
        \centering
        \scalebox{0.9}{
        \begin{tikzpicture}[
                        ]
\node (n1)  [state, initial] {\texttt{0}};
\node (n2)  [state,right=of n1]   {\texttt{1}};
\path   (n1)    edge [loop above] node {\tt a} (n1)
                edge [bend left=15] node {\tt b} (n2)
        (n2)    edge [bend left=15] node {\tt b} (n1)
            edge[loop above] node {\tt a} (n2);
\end{tikzpicture}}
        \caption{Agent~$i$'s MDP in Example~1. Each agent starts in state~\texttt{0} and only has 2 states and 2 actions. Taking action~\texttt{a} in any state will return the same state with probability~$p$ and will transition states with probability~$1-p$. Taking action~\texttt{b} will transition the agent to the other state with probability~$p$ and remain in the same state with probability~$1-p$. 
        }
        \label{fig:toy_MDP}
    \end{figure}

\begin{figure}
        \centering
\begin{tikzpicture}

\definecolor{darkgray176}{RGB}{176,176,176}
\definecolor{darkorange25512714}{RGB}{255,127,14}
\definecolor{steelblue31119180}{RGB}{31,119,180}
\definecolor{gainsboro229}{RGB}{229,229,229}
\definecolor{dimgray85}{RGB}{85,85,85}
\definecolor{UFOrange}{RGB}{250, 70, 22}
\definecolor{UFblue}{RGB}{0, 33, 165}
\definecolor{GTblue}{RGB}{0, 48, 87}
\definecolor{GTgold}{RGB}{179, 163, 105}
\begin{axis}[
height=0.5\figH,
tick align=outside,
tick pos=left,
width=0.35\figW,
axis background/.style={fill=gainsboro229},
axis line style={white},
scale only axis,
x grid style={darkgray176},
xlabel={Number of agents, $N$},
xtick style={color=black},
y grid style={darkgray176},
ylabel={Cost of Privacy},
ytick style={color=black},
yticklabel={$\pgfmathprintnumber{\tick}$\%},
xmajorgrids,
ymajorgrids,
yminorgrids,
yminorticks=true,
y grid style={white},
xtick style={color=dimgray85},
x grid style={white},
legend cell align={left},
legend style={nodes={scale=0.8, transform shape}},
legend style={fill opacity=0.8, draw opacity=1, text opacity=1, draw=white, fill=white},
legend pos = north west,
tick align=outside,
tick pos=left,
xtick={2,3,4, 5, 6}
]
\addplot [ultra thick, UFblue, mark=*, mark options={scale=2, solid}]
table {%
2 14.2232678977612
3 17.0780910908583
4 15.4790284940626
5 19.3862673799264
6 23.58573712694901
};
\addlegendentry{Input Perturbation}
\addplot [ultra thick, UFOrange, dashed, mark=square*,mark options={scale=2, solid}]
table {%
2 46.6510171821443
3 126.217039543591
4 297.946020412267
5 483.19158323732
6 644.5172207953024
};
\addlegendentry{Output Perturbation}
\end{axis}

\end{tikzpicture}
        \caption{The change in value function presented as a percentage, i.e.,~$\frac{|v_{\tilde{\pi}^*}(s_0) - v_{\pi^*}(s_0)|}{|v_{\pi^*}(s_0)|}\cdot 100\%$ in Example~1. We compute a policy with~$500$ sample private reward functions for each~$N\in\{2,\ldots, 6\}$, with~$\epsilon = 1$. Input perturbation is virtually unaffected by the growth in~$N$, while performance degrades with growing~$N$ for output perturbation.}
        \label{fig:value_diff_N_agents}
\end{figure}
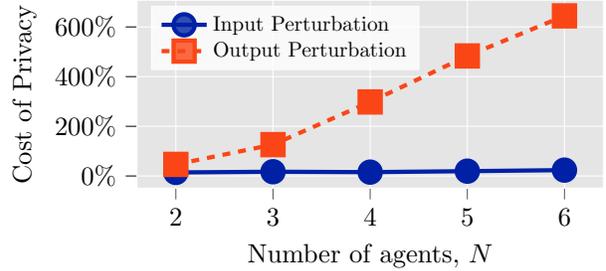

In this example, we consider a team of~$N$ agents operating in an MMDP environment, where each agent is modeled by the MDP shown in Figure~\ref{fig:toy_MDP}. In this case, we have~$\mathcal{S}^i = \{\texttt{0}, \texttt{1}\}$ and~$\mathcal{A}^i = \{\texttt{a}, \texttt{b}\}$ for all~$i\in[N]$. Additionally, each agent has the reward function~$r^i(\texttt{(1, 1,\ldots)}, \texttt{a}) = 5$, and~$r^i(s, a^i) = -1$ for all other state-action pairs. This reward function is designed to drive all agents to remain in state~$\texttt{1}$. We fix privacy parameters~$\epsilon = 1$,~$\delta = 0.1$, and~$b = 2$. We compute the cost of privacy for~$500$ sample private rewards for all~$N\in\{2,\ldots, 6\}.$ The cost of privacy as a function of the number of agents is shown in Figure~\ref{fig:value_diff_N_agents}. As the number of agents increases, input perturbation is virtually unaffected by increasing MMDP size, while the performance of output perturbation degrades substantially as~$N$ increases. The substantially lower cost of privacy for input perturbation motivates our continued use of it below.


\subsection{Example 2: Gridworld}
\label{example:gridworld}
\begin{figure}
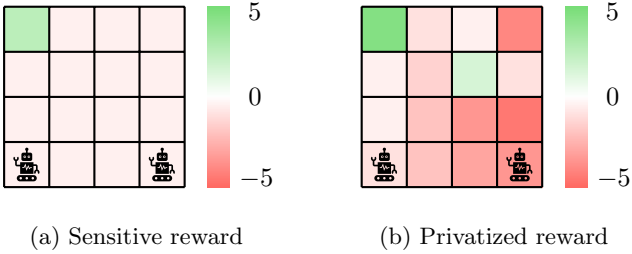

\centering
    \begin{subfigure}{0.2\textwidth}
        \centering
         \input{Tikz_Figures/fourbyfourgridworld}
         \caption{Sensitive reward}
    \label{fig:non_private_rewards}
    \end{subfigure}
   \hspace{1cm}
\begin{subfigure}{0.2\textwidth}
\centering
    \input{Tikz_Figures/fourbyfourgridworld_private}
    \caption{Privatized reward}
    \label{fig:private_rewards}
    \end{subfigure}
\caption{Multi-agent gridworld from Example~\ref{example:gridworld} with (a) non-private rewards and (b) privatized rewards. The color of the cell is determined by the joint reward value for each agent selecting \texttt{stay} in that state, with the opacity proportional to the reward value. Even under privacy, the goal state-action pair remains unchanged in this example. Thus, even when perturbing rewards for privacy, agents' policy will drive them to the same terminal state in this example.}
\label{fig:gridworld}
\end{figure}

Consider the~$4\times 4$ gridworld environment from Figure~\ref{fig:non_private_rewards}, where the goal is for both agents to simultaneously occupy state~$0$, denoted by the green cell in Figure~\ref{fig:non_private_rewards}. The set of local agent actions is~$\mathcal{A}^i = \{\texttt{left}, \texttt{right}, \texttt{up}, \texttt{down}, \texttt{stay}\}$, and we simulate two cases: (i)~$r^i((0, 0), \texttt{stay}) = 5$ and (ii)~$r^i((0, 0), \texttt{stay}) = 50$, with~$r^i(s, a^i) = -1$ for all other state-action pairs in both cases. For any action~$a^i\in\mathcal{A}^i$, the probability of agent $i$ slipping into a state not commanded by the action is~$p=0.1$. We consider the adjacency parameter~$b=2$ and privacy parameters~$\epsilon\in[0.1, 10]$ and~$\delta = 0.1$. Figure~\ref{fig:private_rewards} shows a sample reward function generated with~$\epsilon = 1$. 

Simulating this environment with~$N=2$ agents, Figure~\ref{fig:value_diff_2_agent} shows the change in utility, denoted by a percent change in the value function, with decreasing strength of privacy (quantified by growing~$\epsilon$). We discuss the results of this simulation in four different cases: (i) input perturbation with $\Rmax = 5$, (ii) input perturbation with $\Rmax = 50$, (iii) output perturbation with $\Rmax = 5$, and (iv) output perturbation with $\Rmax = 50$, and then we compare these results. 

\subsubsection{Input Perturbation}

\emph{$\mathbf{\Rmax = 5}$}: With~$\epsilon = 0.1$, the probability that the reward for the state-action pair~$((0, 0), \texttt{stay})$ remains the the maximum reward for each agent is upper bounded by~$0.6261$ according to Theorem~\ref{theorem:reward_function_design}. When $\epsilon = 1$, this probability is upper bounded by~$0.9961$.

In the non-private case,~$11{,}270{,}400$ iterations were used to compute the policy. On average,~$11{,}272{,}256$ iterations were required to compute the policy on the privatized rewards with~$\epsilon = 1.3$, which is a~$0.016\%$ increase. According to Theorem~\ref{thm:e_cost_of_privacy}, we expect at most a~$6.69\%$ increase in the number of iterations. Such an increase is often acceptable in many applications, though the use of differential privacy substantially outperforms the bound, indicating that the increase in computational complexity is negligible. 
We highlight that~$\epsilon=1.3$ on average yields only a~$5\%$ decrease in performance and has a~$0.016\%$ increase computation time, showcasing that even with reasonably strong privacy users will have negligible changes in performance and computation time. Indeed, these results suggest an inherent compatibility between privacy, performance, and computation.

\emph{$\mathbf{\Rmax = 50}$}: According to Theorem~\ref{theorem:reward_function_design}, the probability that the maximum reward state-action pair remains the same is~$0.9969$ with~$\epsilon = 0.1$, and the same probability is~$1$ with~$\epsilon =1$. Thus, we expect a lower cost of privacy on average because the highest reward state-action pair will remain the same after privatization more often with~$\Rmax = 50$ than with~$\Rmax = 5$. For example, with~$\epsilon = 0.1$, we only see a~$1.5\%$ loss in performance on average with~$\Rmax = 50$, while with the same~$\epsilon$ we see a~$108\%$ loss on average with~$\Rmax = 5$. Thus, a reward with a larger gap between its largest and second largest entries can provide significant robustness to privacy.

\subsubsection{Output Perturbation}

\begin{figure}
    \centering
%
%
\definecolor{chocolate2267451}{RGB}{226,74,51}
\definecolor{dimgray85}{RGB}{85,85,85}
\definecolor{gainsboro229}{RGB}{229,229,229}
\definecolor{lightgray204}{RGB}{204,204,204}
\definecolor{steelblue52138189}{RGB}{52,138,189}
\definecolor{black}{RGB}{0, 0, 0}
\definecolor{GTblue}{RGB}{0, 48, 87}
\definecolor{GTgold}{RGB}{179, 163, 105}
\definecolor{UFOrange}{RGB}{250, 70, 22}
\definecolor{UFblue}{RGB}{0, 33, 165}
\definecolor{soft_green}{RGB}{119, 221, 118}
\begin{tikzpicture}

\begin{axis}[%
width=0.35\figW,
height=0.5\figH,
axis background/.style={fill=gainsboro229},
axis line style={white},
scale only axis,
xlabel=\textcolor{black}{{Strength of Privacy, $\epsilon$}},
xtick style={color=dimgray85},
x grid style={white},
yminorticks=true,
y grid style={white},
ylabel=\textcolor{black}{Cost of Privacy},
yticklabel={$\pgfmathprintnumber{\tick}$\%},
xmajorgrids,
ymajorgrids,
yminorgrids,
legend cell align={left},
legend style={fill opacity=0.8, draw opacity=1, text opacity=1, draw=white, fill=white},
legend style={nodes={scale=0.6, transform shape}},
legend pos = north east,
tick align=outside,
tick pos=left
]
\addplot [ultra thick, UFblue, mark=triangle*, mark options={scale=1
, solid}]
table {%
0.1 192.0319484
1 17.47543169
1.1 11.27980098
1.2 7.197752612
1.3 4.993605844
1.4 3.413166678
1.5 4.537233049
2 3.443807184
3 2.457558199
4 2.728050962
5 0.406511205
6 3.139500313
7 2.182596525
8 1.944780564
9 1.814798463
10 2.343884357
};
\addlegendentry{Input Perturbation, $\Rmax = 5$}
\addplot [ultra thick, UFOrange, dashed, mark = triangle*, mark options={scale=1, solid}]
table {%
0.1 270.113371
1 259.0607636
1.5 226.0956488
2 191.6327991
3 120.364221
4 57.87483261
5 29.50706076
6 15.88422343
7 8.306363034
8 6.655503409
9 5.371040826
10 3.951510661
};
\addlegendentry{Output Perturbation, $\Rmax = 5$}
\addplot [ultra thick, purple, mark=x, mark options={scale=1.1}]
table {%
0.1 1.564963732
1 4.594010883
1.2 2.036813234
1.3 0.02106158
1.5 0.075276208
2 0.203498188
3 0.221314551
4 0.003694213
5 0.071385107
6 0.071711455
7 0.255875855
8 0.089616191
9 0.292222615
10 0.170747886
};
\addlegendentry{Input Perturbation, $\Rmax = 50$}
\addplot [ultra thick, soft_green, dashed, mark = x, mark options={scale=1.1, solid}]
table {%
0.1 108.974255
1 2.432587775
1.2 2.38078358
1.3 2.247937887
1.5 2.210385486
2 1.860900394
3 1.558135271
4 1.359591505
5 1.045352777
6 0.896983317
7 0.809722265
8 0.768860103
9 0.655265145
10 0.601462838
};
\addlegendentry{Output Perturbation, $\Rmax = 50$}
\end{axis}
\end{tikzpicture}%
    \caption{The values of~$\frac{|v_{\tilde{\pi}^*}(s_0) - v_{\pi^*}(s_0)|}{| v_{\pi^*}(s_0)|}\cdot 100\%$ for the~$4\times 4$ gridworld in Example~2, averaged over~$1{,}000$ samples at each~$\epsilon$ with~$\epsilon\in[0.1, 10]$ and~$\Rmax = 5$ and~$\Rmax = 50$. Input perturbation has a much lower cost of privacy for all~$\epsilon$ shown. In the case of~$\Rmax = 5$, output perturbation requires significantly weaker privacy, namely~$\epsilon = 6$, to recover the same level of performance that input perturbation has with much stronger privacy, namely~$\epsilon = 1$.}
    \label{fig:value_diff_2_agent}
\end{figure}
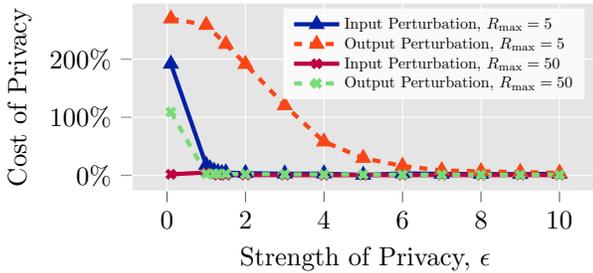

Output perturbation has a greater cost of privacy due to the essentially exponential dependence on the number of agents in calibrating the variance of the noise added to enforce differential privacy with both~$\Rmax = 5$ and~$\Rmax = 50$. The cost of privacy decreases with~$\epsilon$, which intuitively agrees with how privacy should affect the performance of the MMDP (since privacy weakens as~$\epsilon$ grows). The cost of privacy decreases for both input and output perturbation as strength of privacy decreases, implying improved performance as~$\epsilon$ increases. However, agents always perform significantly better when using input perturbation, as shown in Figure~\ref{fig:value_diff_2_agent}.

\subsection{Example 3: Waypoint Guidance}
\label{ex:waypoint}  
\begin{figure}[t]
    \centering
    \begin{subfigure}{0.2\textwidth}
        \centering
         \includegraphics[scale=.75]{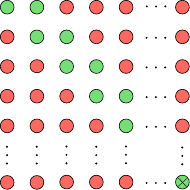}
         \caption{Non-private rewards}
    \label{fig:non_private_waypoints}
    \end{subfigure}
    \begin{subfigure}{0.2\textwidth}
        \centering
    \includegraphics[scale=.75]{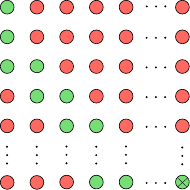}
    \caption{Privatized rewards}
    \label{fig:private_waypoints}
    \end{subfigure}
\caption{Waypoints selected (green) by policy generated using the (a) non-private reward and (b) private reward. In this sample, the goal state, denoted by an ``x", remains unchanged by privacy, however the path generated using the privately generated waypoints is different from that of the nominal set of waypoints, and the number of waypoints required varies case-by-case.}
\label{fig:waypoints}
\end{figure}

     We consider a 3 degree-of-freedom vehicle model, with waypoints that exist in the horizontal plane of the initial condition. The initial position of the vehicle is the first waypoint and the initial state of the MDP. We consider an MDP with a~$20\times 20$ set of waypoints as states, as seen in Figure~\ref{fig:waypoints}, along with the actions
         $\mathcal{A} = \{\texttt{north}, \texttt{south}, \texttt{east}, \texttt{west}, \texttt{to target}\}.$
 Once the vehicle is closer to the new waypoint than the previous one, we say that the new waypoint is the new state of the vehicle, and the policy again commands the vehicle to navigate to a subsequent waypoint. This process continues until the vehicle is less than~$80~\text{km}$ from the target, at which point the vehicle is treated as having reached the target, and is commanded to navigate directly to its location. 
Without this waypoint approach, the vehicles target location may easily be identified by an adversary, and thus we use differential privacy to develop a sequence of waypoints to conceal the vehicles target location until it is within the threshold to disable privacy.
We then implement differential privacy for the reward function of this MDP using Algorithm~\ref{algo:input}. Note that since there is only one agent, the outputs of Algorithm~\ref{algo:input} and Algorithm~\ref{algo:output} are identical. 
    
\begin{figure}[t]
    \centering
    \input{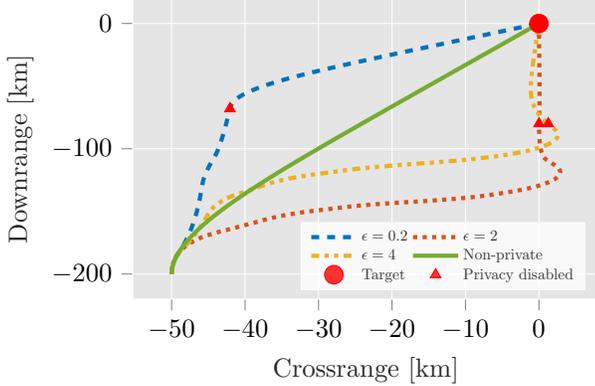}
    \caption{High-speed vehicle trajectories given varying privacy levels from Example~3. 
    The ``Privacy Disabled" marker is placed where the vehicle comes within~$80~\text{km}$ of its target.}
    \label{fig:private_missiles}
\end{figure}

Trajectories generated using various~$\epsilon$ values are presented in Figure~\ref{fig:private_missiles}.
 We simulate~$200$ privatized policies for each~$\epsilon\in[0.2, 4]$. The nominal trajectory uses a generalized vector explicit (GENEX) guidance law~\cite{ohlmeyer2006generalized} to navigate directly to the target from the initial position. We consider an adversary that can measure position, velocity, and acceleration, and knows the initial position of the vehicle, attempting to identify the target location. The adversary assumes a ballistic trajectory from the vehicle, that is, they estimate vehicle's acceleration from~$\ddot{x}(t) = \hat{a}(t_0)e^{-\lambda(t-t_0)},t\in[t_0, t_f]$ where~$\hat{a}$ is the acceleration measurement and~$\lambda$ is a tunable model parameter. The adversary integrates this acceleration estimate to predict the target location of the vehicle. We assume that the adversary knows the final time of the trajectory,~$t_f$. While this would be unrealistic in practice, we note that any realistic adversary would perform worse, and thus these results represent a worst-case bound on adversary performance. We consider the vehicle ``detected" when the adversary's estimated final position,~$x(t_f)$, is within a sufficiently small ball of radius~$\rho$ around the target for three consecutive time steps.

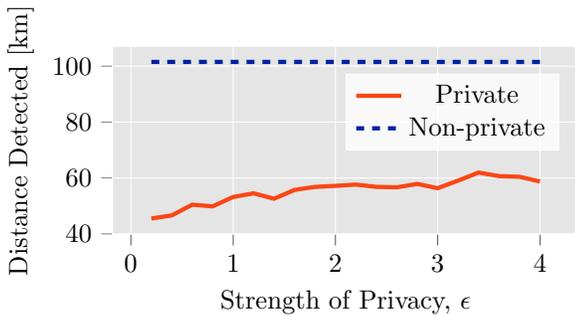
\begin{figure}[h]
    \centering
%
%
\definecolor{chocolate2267451}{RGB}{226,74,51}
\definecolor{dimgray85}{RGB}{85,85,85}
\definecolor{gainsboro229}{RGB}{229,229,229}
\definecolor{lightgray204}{RGB}{204,204,204}
\definecolor{steelblue52138189}{RGB}{52,138,189}
\definecolor{black}{RGB}{0, 0, 0}
\definecolor{UFOrange}{RGB}{250, 70, 22}
\definecolor{UFblue}{RGB}{0, 33, 165}
\definecolor{GTblue}{RGB}{0, 48, 87}
\definecolor{GTgold}{RGB}{179, 163, 105}
\begin{tikzpicture}

\begin{axis}[%
width=0.35\figW,
height=0.5\figH,
axis background/.style={fill=gainsboro229},
axis line style={white},
scale only axis,
xlabel=\textcolor{black}{Strength of Privacy, $\epsilon$},
xtick style={color=dimgray85},
x grid style={white},
yminorticks=true,
y grid style={white},
ylabel=\textcolor{black}{Distance Detected [km]},
xmajorgrids,
ymajorgrids,
yminorgrids,
tick align=outside,
tick pos=left,
legend style={at={(0.5,0.65)},anchor=west},
    legend style={fill opacity=0.8, draw opacity=1, text opacity=1, draw=white, fill=white},
]
\addplot [color=UFOrange, ultra thick]
  table[row sep=crcr]{%
0.2	45.4936898691505\\
0.4	46.6090175230874\\
0.6	50.4433076921836\\
0.8	49.841299189508\\
1	53.1883823437857\\
1.2	54.5112696670394\\
1.4	52.5729469338775\\
1.6	55.7259671261916\\
1.8	56.7976062516562\\
2	57.1758818863801\\
2.2	57.6363416710806\\
2.4	56.8070209885277\\
2.6	56.650369319446\\
2.8	57.8715943211195\\
3	56.3320824790022\\
3.2	59.0562797993236\\
3.4	61.9279828777868\\
3.6	60.6518080815025\\
3.8	60.4216904925985\\
4	58.7214262523086\\
};
\addlegendentry{Private}
    \addplot [mark=none, color = UFblue, ultra thick, , dashed, domain = 0.2:4] {101.5};
    \addlegendentry{Non-private}
\end{axis}
\end{tikzpicture}%
    \caption{Average distance from the target detected for~$\epsilon\in[0.2, 4]$ with~$\rho = 0.7$ km and~$\lambda = 0.1$. As privacy weakens, the vehicle's target is identified further away, and even with weaker privacy, i.e.,~$\epsilon = 4$, the vehicle is undetected after privacy is disabled at~$80$ km. Without privacy, the vehicle is detected at~$101.5$ km, highlighting that even with moderate privacy, the vehicle gets considerably closer to the target before detection.
    }
    \label{fig:deception}
\end{figure}

To assess the impact of privacy, we compare the distance 
from the target at which
the vehicle's intended destination is detected 
both with and without privacy implemented.
These comparisons are shown in Figure~\ref{fig:deception}. With the initial position~$x_0 = [-200~~-50~~10]^T$ km and~$\rho = 0.7$ km, the destination of the non-private trajectory is detected by the adversary~$101.5~\text{km}$ from the target. With strong privacy, i.e.,~$\epsilon = 0.2$, the vehicle gets~$54$ km closer to the target without detection than without privacy. Since privacy is disabled at~$80~\text{km}$, we see that with privacy as weak as~$\epsilon = 4$, on average the vehicle is not detected prior to privacy being disabled.

\section{Conclusion}\label{sec:conclusion}

We have developed two methods for protecting reward functions in MMDPs from observers by using differential privacy, and we identified input perturbation as the more tractable method, which is counter to much of the existing literature. We also developed guidelines for designing reward functions, investigated the additional computational time required to compute a policy on private rewards, and showed the success of these methods in simulation. Future work will include performing experiments with these methods on mobile robots, and investigating the privatization of rewards in a learning setting.
\bibliographystyle{plain}        
\bibliography{main}  

\begin{thebibliography}{10}

\bibitem{arnold1979bounds}
Barry~C Arnold and Richard~A Groeneveld.
\newblock Bounds on expectations of linear systematic statistics based on
  dependent samples.
\newblock {\em The Annals of Statistics}, pages 220--223, 1979.

\bibitem{benvenuti2023differentially}
Alexander Benvenuti, Calvin Hawkins, Brandon Fallin, Bo~Chen, Brendan Bialy,
  Miriam Dennis, and Matthew Hale.
\newblock Differentially private reward functions for markov decision process.
\newblock In {\em 2024 IEEE 8th Conference on Control Technology and
  Applications (CCTA), Accepted}, 2024.

\bibitem{boutilier1996planning}
Craig Boutilier.
\newblock Planning, learning and coordination in multiagent decision processes.
\newblock In {\em TARK}, volume~96, pages 195--210. Citeseer, 1996.

\bibitem{chen2023differential}
Bo~Chen, Calvin Hawkins, Mustafa~O Karabag, Cyrus Neary, Matthew Hale, and Ufuk
  Topcu.
\newblock Differential privacy in cooperative multiagent planning.
\newblock In {\em Uncertainty in Artificial Intelligence}, pages 347--357.
  PMLR, 2023.

\bibitem{chen2023differentialsymbolic}
Bo~Chen, Kevin Leahy, Austin Jones, and Matthew Hale.
\newblock Differential privacy for symbolic systems with application to markov
  chains.
\newblock {\em Automatica}, 152:110908, 2023.

\bibitem{chen2023differentially}
Bo~Chen, Baike She, Calvin Hawkins, Alex Benvenuti, Brandon Fallin, Philip~E
  Par{\'e}, and Matthew Hale.
\newblock Differentially private computation of basic reproduction numbers in
  networked epidemic models.
\newblock In {\em 2024 American control conference (ACC), To Appear}. IEEE,
  2024.

\bibitem{cortes2016differential}
Jorge Cort{\'e}s, Geir~E Dullerud, Shuo Han, Jerome Le~Ny, Sayan Mitra, and
  George~J Pappas.
\newblock Differential privacy in control and network systems.
\newblock In {\em 55th IEEE Conference on Decision and Control (CDC)}, pages
  4252--4272, 2016.

\bibitem{dobbe2018customized}
Roel Dobbe, Ye~Pu, Jingge Zhu, Kannan Ramchandran, and Claire Tomlin.
\newblock Customized local differential privacy for multi-agent distributed
  optimization.
\newblock {\em arXiv preprint arXiv:1806.06035}, 2018.

\bibitem{duchi2013local}
John~C Duchi, Michael~I Jordan, and Martin~J Wainwright.
\newblock Local privacy and statistical minimax rates.
\newblock In {\em 2013 IEEE 54th Annual Symposium on Foundations of Computer
  Science}, pages 429--438. IEEE, 2013.

\bibitem{cynthia2006differential}
Cynthia Dwork.
\newblock Differential privacy.
\newblock {\em Automata, languages and programming}, pages 1--12, 2006.

\bibitem{dwork2014algorithmic}
Cynthia Dwork and Aaron Roth.
\newblock The algorithmic foundations of differential privacy.
\newblock {\em Foundations and Trends in Theoretical Computer Science},
  9(3--4):211--407, 2014.

\bibitem{fallin2023differential}
Brandon Fallin, Calvin Hawkins, Bo~Chen, Parham Gohari, Alexander Benvenuti,
  Ufuk Topcu, and Matthew Hale.
\newblock Differential privacy for stochastic matrices using the matrix
  dirichlet mechanism.
\newblock In {\em 2023 62nd IEEE Conference on Decision and Control (CDC)},
  pages 5067--5072. IEEE, 2023.

\bibitem{glancy2012privacy}
Dorothy~J Glancy.
\newblock Privacy in autonomous vehicles.
\newblock {\em Santa Clara L. Rev.}, 52:1171, 2012.

\bibitem{gohari2020privacy}
Parham Gohari, Matthew Hale, and Ufuk Topcu.
\newblock Privacy-preserving policy synthesis in markov decision processes.
\newblock In {\em 2020 59th IEEE Conference on Decision and Control (CDC)},
  pages 6266--6271. IEEE, 2020.

\bibitem{guan2018privacy}
Zhitao Guan, Guanlin Si, Xiaosong Zhang, Longfei Wu, Nadra Guizani, Xiaojiang
  Du, and Yinglong Ma.
\newblock Privacy-preserving and efficient aggregation based on blockchain for
  power grid communications in smart communities.
\newblock {\em IEEE Communications Magazine}, 56(7):82--88, 2018.

\bibitem{han2018privacy}
Shuo Han and George~J Pappas.
\newblock Privacy in control and dynamical systems.
\newblock {\em Annual Review of Control, Robotics, and Autonomous Systems},
  1:309--332, 2018.

\bibitem{han2016differentially}
Shuo Han, Ufuk Topcu, and George~J Pappas.
\newblock Differentially private distributed constrained optimization.
\newblock {\em IEEE Transactions on Automatic Control}, 62(1):50--64, 2016.

\bibitem{hassan2021privacy}
Ali Hassan, Deepjyoti Deka, and Yury Dvorkin.
\newblock Privacy-aware load ensemble control: A linearly-solvable mdp
  approach.
\newblock {\em IEEE Transactions on Smart Grid}, 13(1):255--267, 2021.

\bibitem{hawkins2023node}
Calvin Hawkins, Bo~Chen, Kasra Yazdani, and Matthew Hale.
\newblock Node and edge differential privacy for graph laplacian spectra:
  Mechanisms and scaling laws.
\newblock {\em IEEE Transactions on Network Science and Engineering}, 2023.

\bibitem{hawkins2020differentially}
Calvin Hawkins and Matthew Hale.
\newblock Differentially private formation control.
\newblock In {\em 2020 59th IEEE Conference on Decision and Control (CDC)},
  pages 6260--6265. IEEE, 2020.

\bibitem{hsu2014differential}
Justin Hsu, Marco Gaboardi, Andreas Haeberlen, Sanjeev Khanna, Arjun Narayan,
  Benjamin~C Pierce, and Aaron Roth.
\newblock Differential privacy: An economic method for choosing epsilon.
\newblock In {\em 2014 IEEE 27th Computer Security Foundations Symposium},
  pages 398--410. IEEE, 2014.

\bibitem{huang2015differentially}
Zhenqi Huang, Sayan Mitra, and Nitin Vaidya.
\newblock Differentially private distributed optimization.
\newblock In {\em Proceedings of the 16th International Conference on
  Distributed Computing and Networking}, pages 1--10, 2015.

\bibitem{keizer2013training}
Simon Keizer, Mary~Ellen Foster, Oliver Lemon, Andre Gaschler, and Manuel
  Giuliani.
\newblock Training and evaluation of an mdp model for social multi-user
  human-robot interaction.
\newblock In {\em Proceedings of the SIGDIAL 2013 Conference}, pages 223--232,
  2013.

\bibitem{le2013differentially}
Jerome Le~Ny and George~J Pappas.
\newblock Differentially private filtering.
\newblock {\em IEEE Transactions on Automatic Control}, 59(2):341--354, 2013.

\bibitem{leone1961folded}
Fred~C Leone, Lloyd~S Nelson, and RB~Nottingham.
\newblock The folded normal distribution.
\newblock {\em Technometrics}, 3(4):543--550, 1961.

\bibitem{lv2020differentially}
Yuan-Wei Lv, Guang-Hong Yang, and Chong-Xiao Shi.
\newblock Differentially private distributed optimization for multi-agent
  systems via the augmented lagrangian algorithm.
\newblock {\em Information Sciences}, 538:39--53, 2020.

\bibitem{ma2019differentially}
Pingchuan Ma, Zhiqiang Wang, Le~Zhang, Ruming Wang, Xiaoxiang Zou, and Tao
  Yang.
\newblock Differentially private reinforcement learning.
\newblock In {\em International Conference on Information and Communications
  Security}, pages 668--683. Springer, 2019.

\bibitem{ng2000algorithms}
Andrew~Y Ng, Stuart Russell, et~al.
\newblock Algorithms for inverse reinforcement learning.
\newblock In {\em Icml}, volume~1, page~2, 2000.

\bibitem{nozari2016differentially}
Erfan Nozari, Pavankumar Tallapragada, and Jorge Cort{\'e}s.
\newblock Differentially private distributed convex optimization via functional
  perturbation.
\newblock {\em IEEE Transactions on Control of Network Systems}, 5(1):395--408,
  2016.

\bibitem{ohlmeyer2006generalized}
Ernest~J Ohlmeyer and Craig~A Phillips.
\newblock Generalized vector explicit guidance.
\newblock {\em Journal of guidance, control, and dynamics}, 29(2):261--268,
  2006.

\bibitem{puterman2014markov}
Martin~L Puterman.
\newblock {\em Markov decision processes: discrete stochastic dynamic
  programming}.
\newblock John Wiley \& Sons, 2014.

\bibitem{ramirez2011goal}
Miquel Ram{\i}rez and Hector Geffner.
\newblock Goal recognition over pomdps: Inferring the intention of a pomdp
  agent.
\newblock In {\em IJCAI}, pages 2009--2014. IJCAI/AAAI, 2011.

\bibitem{venkitasubramaniam2013privacy}
Parv Venkitasubramaniam.
\newblock Privacy in stochastic control: A markov decision process perspective.
\newblock In {\em 51st Annual Allerton Conference on Communication, Control,
  and Computing}, pages 381--388, 2013.

\bibitem{wachi2020safe}
Akifumi Wachi and Yanan Sui.
\newblock Safe reinforcement learning in constrained markov decision processes.
\newblock In {\em International Conference on Machine Learning}, pages
  9797--9806. PMLR, 2020.

\bibitem{wang2016differentially}
Yu~Wang, Matthew Hale, Magnus Egerstedt, and Geir~E Dullerud.
\newblock Differentially private objective functions in distributed cloud-based
  optimization.
\newblock In {\em 2016 IEEE 55th Conference on Decision and Control (CDC)},
  pages 3688--3694. IEEE, 2016.

\bibitem{yazdani2022differentially}
Kasra Yazdani, Austin Jones, Kevin Leahy, and Matthew Hale.
\newblock Differentially private lq control.
\newblock {\em IEEE Transactions on Automatic Control}, 2022.

\bibitem{ye2019differentially}
Dayong Ye, Tianqing Zhu, Wanlei Zhou, and S~Yu Philip.
\newblock Differentially private malicious agent avoidance in multiagent
  advising learning.
\newblock {\em IEEE transactions on cybernetics}, 50(10):4214--4227, 2019.

\bibitem{zhi2020online}
Tan Zhi-Xuan, Jordyn Mann, Tom Silver, Josh Tenenbaum, and Vikash Mansinghka.
\newblock Online bayesian goal inference for boundedly rational planning
  agents.
\newblock {\em Advances in neural information processing systems},
  33:19238--19250, 2020.

\bibitem{zhou2022differentially}
Xingyu Zhou.
\newblock Differentially private reinforcement learning with linear function
  approximation.
\newblock {\em Proceedings of the ACM on Measurement and Analysis of Computing
  Systems}, 6(1):1--27, 2022.

\bibitem{ziebart2008maximum}
Brian~D Ziebart, Andrew~L Maas, J~Andrew Bagnell, Anind~K Dey, et~al.
\newblock Maximum entropy inverse reinforcement learning.
\newblock In {\em Aaai}, volume~8, pages 1433--1438. Chicago, IL, USA, 2008.

\end{thebibliography}
\begin{appendices}
    \section{Proof of Theorem~\ref{thm:output}}\label{apdx:thm_2}
From Lemma~\ref{lem:gauss_mech}, Algorithm~\ref{algo:output} is differentially private if~$\sigma$ is chosen according to 
    $\sigma\geq \frac{\Delta_2 \rew}{2\epsilon}\kappa(\epsilon, \delta).$
We now compute~$\Delta_2 \rew$ and substitute it the lower bound on~$\sigma$, which will complete the proof, since the computation of~$\tilde{\pi}^{*, i}$ is differentially private in accordance with Lemma~\ref{lem:arbitrary}.
For each~$i\in[N]$, let~$r^i$ and~$\hat{r}^{i}$ denote two adjacent reward functions and let~$R^i$ and~$\hat{R}^{i}$ denote their vectorized forms as defined in~\eqref{eq:big_ri}. Then
\begin{multline}
    R = \frac{1}{N}\Big[
    \sum_{i\in[N]} r^i\left(s_1, a^i_{I_1(i)}\right),
    \cdots,
    \sum_{i\in[N]} r^i\left(s_1, a^i_{I_m(i)}\right),\\
    \sum_{i\in[N]} r^i\left(s_2, a^i_{I_1(i)}\right),
    \cdots
    \sum_{i\in[N]} r^i\left(s_{n}, a^i_{I_m(i)}\right)
    \Big]^T\in\mathbb{R}^{nm},
\end{multline}
with~$\hat{R}$ defined analogously. 
As noted in Remark~\ref{remark:output},~$R$ and~$\hat{R}$ can differ for only one agent. Suppose the index of that agent is~$j$.
We then determine the sensitivity of~$\rew$ from Definition \ref{def:sensitivity} as follows:
    $\Delta_2 \rew = \max_{\substack{\Adjacent{R^j}{\hat{R}^j};R^i=\hat{R}^i, i\neq j}} \| R - \hat{R}\|_2.$
Since~$R^j$ and~$\hat{R}^{j}$ are~$b$-adjacent, there exists some indices~$k \in [n]$ and~$\ell \in [m_j]$ such that~$r^j(s_k, a^j_{\ell}) \neq \hat{r}^{j}(s_k, a^j_{\ell})$ and~$r^j(s_c, a^j_d) = \hat{r}^{j}(s_c, a^j_d)$ for 
all~$c \in [n] \backslash \{k\}$ and~$d \in [m_j] \backslash \{\ell\}$. 
Since~$R^j$ and~$\hat{R}^{j}$ are~$b$-adjacent, we obtain
\begin{multline}\label{eq:badjacent}
    \Delta_2 \rew = \max_{\substack{\Adjacent{R^j}{\hat{R}^j}\\R^i=\hat{R}^i, i\neq j}} \frac{1}{N}\Big\lVert\Big[
        0,
        \cdots,
        r^j(s_k, a^j_{\ell}) - \hat{r}^{j}(s_k, a^j_{\ell}),\\
        \cdots,
        r^j(s_k, a^j_{\ell}) - \hat{r}^{j}(s_k, a^j_{\ell}),
        \cdots,
        0,
        \cdots,
        0
    \Big]^T\Big\rVert_2,
\end{multline}
where~$r^j(s_k, a^j_{\ell}) - \hat{r}^{j}(s_k, a^j_{\ell})$ appears~$\prod_{\substack{\ell = 1\\ \ell \neq j}}^N m_{\ell}$ times since the local action~$a^j_{\ell}$ appears in~$\prod_{\substack{\ell = 1\\ \ell \neq j}}^N m_{\ell}$ joint actions. 

Since~$\|R-\hat{R}\|_2\leq\|R-\hat{R}\|_1$, we upper bound \eqref{eq:badjacent} as 
   $\Delta_2 \rew \leq \frac{1}{N}|r^j(s_k, a^j_{\ell}) - \hat{r}^{j}(s_k, a^j_{\ell})|\prod_{\substack{\ell = 1\\ \ell \neq j}}^N m_{\ell}.$
From Definition~\ref{def:adj2},~$|r^j(s_k, a^j_{\ell}) - \hat{r}^{j}(s_k, a^j_{\ell})| \leq b$, which we use 
to find
    $\Delta_2 \rew \leq \frac{b}{N}\max_{j\in[n]}\prod_{\substack{\ell = 1\\ \ell \neq j}}^N m_\ell = \frac{b}{N}\productthing,$
which we substitute into~$\frac{\Delta_2 \rew}{2\epsilon}\kappa(\epsilon, \delta)$ to complete the proof.\hfill$\blacksquare$

\section{Proof of Theorem \ref{theorem:cheb}}\label{apx:thm_3}
First we analyze the vectorized rewards~$R$ and~$\tilde{R}$. For each~$i\in[N]$, since~$\tilde{r}^i$ is generated using Algorithm~\ref{algo:input}, we have~$\tilde{r}^i(s_k, a^i_{I_j(i)}) = r^i(s_k, a^i_{I_j(i)})+w^k_{I_j(i)}$, where~$w^k_{I_j(i)}\sim\mathcal{N}(0, \sigma^2)$. For any~$s_k\in\mathcal{S}$ and~$a_j\in\mathcal{A}$ the corresponding entry of~$\tilde{R} - R$ is
     $\frac{1}{N} \sum_{i\in[N]} \tilde{r}^i(s_k, a^i_{I_j(i)}) -  r^i(s_k, a^i_{I_j(i)}) = \frac{1}{N} \sum_{i\in[N]} w^k_{I_j(i)}.$
Thus, the entirety of~$\tilde{R}-R$ is  
\begin{multline}\label{eq:sumofnnoise_cheb}
    \tilde{R}-R=\frac{1}{N}\Big[
    \sum_{i\in[N]} w^1_{I_1(i)},
    \sum_{i\in[N]} w^1_{I_2(i)},
    \cdots, \\
    \sum_{i\in[N]} w^2_{I_1(i)},
    \sum_{i\in[N]} w^2_{I_2(i)},
    \cdots,
    \sum_{i\in[N]} w^n_{I_m(i)}
    \Big]^T.
    \end{multline}
    Let $y_{\ell}$ be the $\ell^{th}$ entry of $\tilde{R}-R$.
    Now the maximum expected absolute error over the entries is
       $\Expectation{\max_{\ell} \left|y_{\ell}\right|}$.
   Using~\eqref{eq:sumofnnoise_cheb} and the fact that~$w^k_{I_j(i)}\sim\mathcal{N}(0, \sigma^2)$ for all~$i, j,$~and~$k$, we see that for any $\ell$ we have $y_{\ell}\sim\mathcal{N}\big(0, \frac{\sigma^2}{N}\big)$. Thus, $\left|y_{\ell}\right|$ has a ``folded-normal" distribution~\cite{leone1961folded} because it is equal to the absolute value of a normal random variable. 
   As a result, 
       $\Expectation{\left|y_{\ell}\right|} = \sigma\sqrt{\frac{2}{N\pi}}, \quad \Var{\left|y_{\ell}\right|} = \frac{\sigma^2}{N}\left(1-\frac{2}{\pi}\right).$
   Then using \cite[Eq.(4)]{arnold1979bounds} for~$|y_{\ell}|$ with $k=nm$ gives
       $\Expectation{\max_{\ell} \left|y_{\ell}\right|} \leq\Expectation{\left|y_{\ell}\right|}+\sqrt{\Var{|y_{\ell}|}(nm-1)}$. 
   Expanding the expectation and variance of~$|y_{\ell}|$ and substituting in $\sigma = \frac{b}{2\epsilon}\kappa(\epsilon, \delta)$ gives the result. \hfill$\blacksquare$

   \section{Proof of Corollary \ref{corollary:choose_epsilon}}\label{apdx:cor1}
   Ensuring that the upper bound from Theorem~\ref{theorem:cheb} is less than or equal to~$A$ provides a sufficient condition for~$\Expectation{\max_{k,j}|\tilde{r}(s_k, a_j)-r(s_k, a_j)|}\leq A.$ We solve for a condition on~$\epsilon$ by rearranging~\eqref{eq:accuracy_result} to find
        $\frac{Cb}{2\epsilon}( \mathcal{Q}^{-1}(\delta)+\sqrt{\mathcal{Q}^{-1}(\delta)^2+2\epsilon} )\leq A$, which we rearrange to obtain
        $\sqrt{\mathcal{Q}^{-1}(\delta)^2+2\epsilon} \leq \frac{2A\epsilon}{Cb}-\mathcal{Q}^{-1}(\delta).$
    Squaring both sides and expanding gives
        $\frac{4A^2}{C^2b^2}\epsilon^2-(\frac{4A\mathcal{Q}^{-1}(\delta)}{Cb}+2)\epsilon=
        \epsilon(\frac{4A^2}{C^2b^2}\epsilon-\frac{4A\mathcal{Q}^{-1}(\delta)}{Cb}-2)$,
    where both sides are non-negative if
        $\frac{4A^2}{C^2b^2}\epsilon-\frac{4A\mathcal{Q}^{-1}(\delta)}{Cb}-2 \geq 0.$
    Solving for~$\epsilon$ completes the proof. \hfill$\blacksquare$

\section{Proof of Theorem \ref{theorem:cop}}\label{adpx:thm_4}
    First, for the cost of privacy $|v_{\tilde{\pi}^*}(s_0) - v_{\pi^*}(s_0)|$ to be within~$\eta$ of its exact value, it is sufficient to compute both~$v_{\tilde{\pi}^*}$ and ~$v_{\pi^*}$ each to within~$\frac{\eta}{2}$ of their exact values.
Next, we define $R_{\text{max}} = \max_{(s, a)\in\mathcal{S}\times\mathcal{A}}|r(s, a)|$ and $\tilde{R}_{\text{max}} = \max_{(s, a)\in\mathcal{S}\times\mathcal{A}}|\tilde{r}(s, a)|$.
Now suppose that value iteration is used to compute~$\tilde{\pi}$. 
Consider the vectorized value function at convergence, namely $V_{\pi^*}$, and the vectorized value function at iteration $k+1$, namely $V_{k+1}$. 
From Proposition \ref{prop:contract}, we have 
    $\norm{V_{k+1}-V_{\pi^*}}_{\infty} = \norm{\mathcal{L}V_k-\mathcal{L}V_{\pi^*}}_{\infty} \leq \gamma\norm{V_{k}-V_{\pi^*}}_{\infty}$,
giving~$\norm{V_{k+1} \!-\! V_{\pi^*}}_{\infty} \!\leq\! \frac{\gamma^k}{1 - \gamma}\norm{V^{1}_{\pi^*} \!-\! V^{0}_{\pi^*}}_{\infty}$ and thus
\begin{equation}\label{eq:upper_bound_this}
    \frac{\gamma^k}{1 \!-\! \gamma}\norm{V^{1}_{\pi^*} \!-\! V^{0}_{\pi^*}}_{\infty} \leq 2\gamma^k \max\limits_{s\in\mathcal{S}}V_{\pi^*}(s)\frac{1}{1 - \gamma},
\end{equation}
where~$V^0_{\pi^*}$ and~$V^1_{\pi^*}$ are, respectively, the value functions vectors attained after~$0$ and~$1$ iterations of the policy evaluation. However,~$\max_{s\in\mathcal{S}}V_{\pi^*}(s)$ may not be known exactly, so we upper bound~\eqref{eq:upper_bound_this} using~$\max_{s\in\mathcal{S}}V_{\pi^*}(s)\leq \frac{R_{\text{max}}}{1-\gamma}$ to find
   $\frac{2\gamma^k\max_{s\in\mathcal{S}}V_{\pi^*}(s)}{1-\gamma} \leq \frac{2\gamma^kR_{\text{max}}}{(1-\gamma)^2}$.
To compute the non-private value function $V_{\pi^*}$ within $\frac{\eta}{2}$ of the limiting value, we wish to find the minimum number of iterations of policy evaluation such that
    $\frac{2\gamma^kR_{\text{max}}}{(1-\gamma)^2}\leq \frac{\eta}{2},$
which we denote by $K_1$. Rearranging, we find
     $K_1 = \ceil{\log(\frac{4R_{\text{max}}}{\eta(1-\gamma)^2})/\log(\frac{1}{\gamma})}.$
Following an identical construction, for computing the private value function vector~$V_{\tilde{\pi}^*}$ to within~$\frac{\eta}{2}$ of its exact value we have that~$K_2 = \ceil{\log(\frac{4\tilde{R}_{\text{max}}}{\eta(1-\gamma)^2})/\log(\frac{1}{\gamma})}$
iterations of policy evaluation are required.
Therefore, the total number of iterations of policy evaluation required to determine the cost of privacy is~$K = K_1+K_2$. 
Additionally, each iteration of policy evaluation requires looping through all~$nm$ values of the reward function. As a result, we require~$nm(K_1+K_2)$ computations to compute the cost of privacy to within~$\eta$ of its exact value, as desired. \hfill$\blacksquare$

\section{Proof of Theorem \ref{theorem:reward_function_design}}\label{adpx:thm_5}
    An upper bound on the probability of the intersection of events is the smallest probability of the individual events. Therefore, we begin by bounding~$\mathbb{P}(B^p \cap B_q) \leq \min\left\{\mathbb{P}(B^p), \mathbb{P}(B_q)\right\}.$
    Thus, we must compute~$\mathbb{P}(B^p)$ and~$\mathbb{P}(B_q)$.
For~$\mathbb{P}(B^p)$, we have a probability of the intersection of dependent events, which is upper bounded by the smallest probability of an event in the intersection. The smallest element in~$T^p(R^i)$ has the smallest probability of remaining larger than every element in~$\tilde{T}^{-p}(R^i)$, which leads to the bound
    \begin{equation}\label{eq:break_up_intersection}
    \!\!\!\mathbb{P}\left(B^p  \right)
    \leq \mathbb{P}\Big(\bigcap_ {k\in[p^-]}\!\!\left[\min \tilde{T}^p(R^i)
    \geq \tilde{T}^{-p}(R^i)_{k}\right]\Big).
    \end{equation}
    We once again have an intersection of dependent events, which we can upper bound using the event with the smallest probability of occurring. The entry closest in value to~$\min T^p(R^i)$, namely~$\max T^{-p}(R^i)$, has the lowest probability of remaining smaller than every element in~$\min T^p(R^i)$ after privatization, so we bound~\eqref{eq:break_up_intersection} by
    \begin{multline}
    \label{eq:this_this_gaussian}
        \mathbb{P}\Big(\bigcap_ {k\in[p^-]}\left[\min \tilde{T}^p(R^i)\geq \tilde{T}^{-p}(R^i)_{k}\right]\Big) \\ 
        \leq \mathbb{P}\left(\tilde{T}^p(R^i)_\text{min}\geq \tilde{T}^{-p}(R^i)_\text{max}\right).
    \end{multline}
     where~$\tilde{T}^p(R^i)_\text{min}$ is~$\min T^p(R^i)$ after privatization and $\tilde{T}^{-p}(R^i)_\text{max}$ is~$\max T^{-p}(R^i)$ after privatization. Since~$\tilde{T}^p(R^i)_\text{min}\sim\mathcal{N}(\min T^p(R^i), \sigma^2)$, with~$\sigma = \frac{b}{2\epsilon}\kappa(\epsilon, \delta)$, the latter event in~\eqref{eq:this_this_gaussian} has a Gaussian distribution, and we write~\eqref{eq:this_this_gaussian} as
        $\mathbb{P}(\tilde{T}^p(R^i)_\text{min}\geq  \tilde{T}^{-p}(R^i)_\text{max})\!\! = \Phi(\frac{\min T^p(R^i) -\max T^{-p}(R^i)}{\sqrt{2}\sigma}).$
    Then we bound~$\mathbb{P}\left(B^p  \right)$ by:
    \begin{equation}\label{eq:p_bound}
    \mathbb{P}\left(B^p  \right) \leq \Phi\Big(\frac{\min T^p(R^i) - \max T^{-p}(R^i)}{\sqrt{2}\sigma} \Big).
    \end{equation}
    The bound on~$\mathbb{P}(B_q)$ follows a similar construction, where we have an intersection of dependent events and thus
    ${\mathbb{P}\left(B_q  \right) \leq \mathbb{P}\big(\bigcap_ {k\in[nm-q]}\big[\max \tilde{T}_q(R^i)\leq \tilde{T}_{-q}(R^i)_{k}\big]\big).}$
    This is once again an intersection of dependent events. Similar to 
    $\mathbb{P}(B^p)$, we upper bound this by the smallest event in this intersection, i.e.,
    \begin{multline}
    \label{eq:min_event}
         \mathbb{P}\Big(\bigcap_ {k\in[q^-]}\left[\max \tilde{T}_q(R^i)\leq \tilde{T}_{-q}(R^i)_{k}\right]\Big)\\ \leq  \mathbb{P}\left(\tilde{T}_q(R^i)_{\text{max}}\leq   \tilde{T}_{-q}(R^i)_{\text{min}}\right).
    \end{multline}
    where~$\tilde{T}_{q}(R^i)_{\text{max}}$ is~$\max T_{q}(R^i)$ after privatization and $\tilde{T}_{-q}(R^i)_{\text{min}}$ is~$\min T_{-q}(R^i)$ after privatization. Since~$\tilde{T}^p(R^i)_\text{min}\sim\mathcal{N}(\max T_q(R^i), \sigma^2)$, the event in~\eqref{eq:min_event} has a Gaussian distribution, from which it follows that
        $\mathbb{P}(\tilde{T}_q(R^i)_{\text{max}}\leq  \tilde{T}_{-q}(R^i)_{\text{min}})\! = \!\mathcal{Q}(\frac{ \max T_q(R^i) - \min T_{-q}(R^i)}{\sqrt{2}\sigma} ).$
    Accordingly, we complete the bound on~$\mathbb{P}\left(B_q  \right)$:
    \begin{equation}\label{eq:q_bound}
        \mathbb{P}\left(B_q  \right)\leq \mathcal{Q}\Big(\frac{\max{T_q(R^i)} - \min{T_{-q}(R^i)}}{\sqrt{2}\sigma} \Big).
    \end{equation}
    Substituting~\eqref{eq:p_bound} and~\eqref{eq:q_bound} into~$\min\left\{\mathbb{P}(B^p), \mathbb{P}(B_q)\right\}$ gives the result. \hfill$\blacksquare$

    \section{Proof of Theorem \ref{thm:e_cost_of_privacy}}\label{apdx:thm_6}
    First, by linearity of the expectation we note that
    $\Expectation{C}=\Expectation{nm(K_1-K_2)}
    =nm(\Expectation{K_1}-K_2).$
We apply the bound~$\ceil*{x}\leq x+1$ and monotinicity of the expectation to get
\begin{equation}\label{eq:ceil_inequality}
    \!\!\!\!nm\left(\Expectation{K_1} \!- \!K_2\right)
   \! \leq 
    nm\left(\frac{\Expectation{\log\left(\frac{4\tilde{R}_{\text{max}}}{\eta(1-\gamma)^2}\right)}}{\log\left(\frac{1}{\gamma}\right)}\!+\!1\!-\!K_2\!\right).
\end{equation}
We will focus on bounding~$\mathbb{E}[\log(\frac{4\tilde{R}_{\text{max}}}{\eta(1-\gamma)^2})]$ and substitute back into~\eqref{eq:ceil_inequality} to complete the proof. Applying Jensen's inequality to $\mathbb{E}[{\log(\frac{4\tilde{R}_{\text{max}}}{\eta(1-\gamma)^2})}]$ gives
\begin{equation}\label{eq:jensen_inequality}
    \mathbb{E}\Big[\log\Big(\frac{4\tilde{R}_{\text{max}}}{\eta(1-\gamma)^2}\Big)\Big]\leq \log\Big(\frac{4\mathbb{E}[\tilde{R}_{\text{max}}]}{\eta(1-\gamma)^2}\Big)\frac{1}{\log\left(\frac{1}{\gamma}\right)}
\end{equation}
Recall that~$\tilde{R}_{\text{max}} = \max_{s, a}|\tilde{r}(s, a)| = \max_{j}|\tilde{R}_j|$, where~$\tilde{R}_j\sim\mathcal{N}(R_{\ell}, \frac{\sigma^2}{N})$. Simply substituting~$\tRmax$ would make it impossible to analytically compare across varying strengths of privacy since doing so would require us to know the maximum reward after privacy is applied, which is information that we do not have access to. To develop a bound independent of privacy, we substitute
    $\max_j|\tilde{R}_j|\leq \max_k |R_k| +\max_j|z_j|,$
into~\eqref{eq:jensen_inequality}, where~$z_i\sim\mathcal{N}(0, \frac{\sigma^2}{N})$ is the perturbation to~$R^i$, to obtain~$\mathbb{E}[\tilde{R}_{\text{max}}]\leq R_{\text{max}}+\Expectation{\max_j|z_j|}$.
It follows that~$|z|$ follows a folded-normal distribution~\cite{leone1961folded}, and accordingly has
$\Expectation{|z|} = \sigma\sqrt{2/(N\pi)}, \Var{|z|} = \frac{\sigma^2}{N}\left(1-\frac{2}{\pi}\right).$
Since~$\max_j|z_j|$ is the first order statistic of the collection~$\{|z_1|,\ldots, |z_{nm}|\}$, we apply~\cite[Eq. (4)]{arnold1979bounds} to obtain
    $\Expectation{\max_j|z_j|} \leq \Expectation{\left|z_{j}\right|} 
       +\sqrt{\Var{\left|z_{j}\right|}(nm-1)}$,
which, after expanding, we substitute into~\eqref{eq:jensen_inequality} and include a ceiling function (since we require this value to be an integer) to complete the proof.\hfill$\blacksquare$
\end{appendices}


\end{document}